\documentclass[aps,prx,groupedaddress,reprint,showpacs,pdftex,superscriptaddress]{revtex4-1}
\usepackage{graphicx}
\usepackage{tabularx}
\usepackage{multirow}
\usepackage{longtable}
\usepackage{dcolumn}
\usepackage{bm}
\usepackage{amsmath}
\usepackage{amssymb}
\usepackage{longtable}
\usepackage{txfonts}
\usepackage{url}
\usepackage[colorlinks=true, linkcolor=blue, anchorcolor=blue, citecolor=blue, urlcolor=magenta]{hyperref}
\usepackage[usenames,dvipsnames]{color}
\definecolor{Gray}{gray}{0.0}
\definecolor{lightGray}{gray}{0.35}

\begin{document}
\title{
Potential high-$T_\mathrm{c}$ superconductivity in YCeH$_{x}$ and LaCeH$_{x}$  under pressure
}
\author{Peng Song}
\affiliation{School of Information Science, JAIST, Asahidai 1-1, Nomi, Ishikawa 923-1292, Japan}
\author{Zhufeng Hou}
\affiliation{State Key Laboratory of Structural Chemistry, Fujian Institute of Research on the Structure of Matter, Chinese Academy of Sciences, Fuzhou 350002, China}
\author{Kousuke Nakano}
\affiliation{School of Information Science, JAIST, Asahidai 1-1, Nomi, Ishikawa 923-1292, Japan}
\affiliation{International School for Advanced Studies (SISSA), Via Bonomea 265, 34136, Trieste, Italy}
\author{Kenta Hongo}
\affiliation{Research Center for Advanced Computing Infrastructure, JAIST, Asahidai 1-1, Nomi, Ishikawa 923-1292, Japan}
\author{Ryo Maezono}
\affiliation{School of Information Science, JAIST, Asahidai 1-1, Nomi, Ishikawa 923-1292, Japan}

\vspace{10mm}

\date{\today}
\begin{abstract}
Lanthanum, yttrium, and cerium hydrides are the three most well-known superconducting binary hydrides
(La-H, Y-H, and Ce-H systems),
which have gained great attention in both theoretical and experimental studies. Recent studies have shown that ternary hydrides composed of lanthanum and yttrium can achieve high superconductivity around 253 K.
In this study we employ the evolutionary-algorithm-based crystal structure prediction (CSP) method and first-principles calculations to investigate the stability and superconductivity of ternary hydrides composed of (Y, Ce) and (La, Ce) under high pressure. Our calculations show that there are multiple stable phases in Y-Ce-H and La-Ce-H systems, among which $P4/mmm$-YCeH$_{8}$, $P\bar{6}m2$-YCeH$_{18}$, $R\bar{3}m$-YCeH$_{20}$, $P4/mmm$-LaCeH$_{8}$, and $R\bar{3}m$-LaCeH$_{20}$ possessing H$_{18}$, H$_{29}$ and H$_{32}$ clathrate structures can maintain both the thermodynamic and lattice-dynamic stabilities.
In addition, we also find that these phases also maintain a strong resistance to decomposition at high temperature.
Electron-phonon coupling calculations show that only three of these five phases can exhibit high-temperature superconductivity.
The superconducting transition temperatures ($T_\mathrm{c}$) of $R\bar{3}m$-YCeH$_{20}$, $R\bar{3}m$-LaCeH$_{20}$, and $P\bar{6}m2$-YCeH$_{18}$ are predicted using the Allen-Dynes-modified McMillan formula to be 122 K at 300 GPa, 116 K at 250 GPa, and 173 K at 150 GPa, respectively. Moreover, the pressure to stabilize  $P\bar{6}m2$-YCeH$_{18}$ can be lowered to 150 GPa, suggesting an accessible condition for its high-pressure synthesis.
\end{abstract}
\maketitle

\section{Introduction}
\label{sec.intro}
Superconductivity has gotten a remarkable
progress and attention owing to its unique phenomenon
and wide range of industrial applications.~\cite{2004MIC,2018GEN,2015LIN,1962JOS,2019BER,2014ISL,2016THO}
On the other hand, the extremely low temperature
environment required for the manifestation of
these properties severely limits the breadth of their application.
Therefore, the pursuit for high(room)-temperature superconductivity
has historically been one of the most competitive and sought goals in superconductivity.~\cite{1988TIN,2011KAL,2020FLO}
According to the Bardeen-Cooper-Schrieffer (BCS) theory,
the superconducting transition temperature ($T\rm_{c}$) significantly correlates with
the phonon vibration frequency.~\cite{1955BAR,1956COO,1957BARa,1957BARb}
In this sense, metallic hydrogen
is considered to be the most ideal candidate
for high-$T\rm_{c}$ superconductivity due to its ultra-high phonon vibration frequency.~\cite{1968ASH}
However, hydrogen metallization, also known as the Wigner-Huntington transition,
typically requires extremely high pressures, making it difficult to
study its superconductivity-related electrical properties at these pressures.~\cite{1935WIG,1996LOU,2015MCM,2011ERE,2016DAL,2017DIA,2017CAS,2018CEL}
Ashcroft proposed the introduction of other elements into hydrogen can
provide the necessary pre-compression for the entire system, allowing it
to maintain metallicity and superconductivity at lower pressures.~\cite{2004ASH}
Thus, metal hydrides are one of the most ideal candidates for high-$T\rm_{c}$ superconductivity.

\vspace{2mm}
According to Ashcroft, almost all binary hydrides have been theoretically screened out by density functional theory (DFT) calculations, and some predictions have been verified by experiments.~\cite{2020SEM,2017PEN,2020ZHA,2017LI,2015YU,2016EIN,2015DRO,2019DRO,2019SAL,2021CHEa,2021SHAa,2021SHAb,2021TRO,2021CHEb,2021KON,2014LI,2019SOM}
Compared to binary superconducting hydrides, research on multi-component
superconducting hydrides appears to be slightly less.
Since multi-component hydrides contain more combinations of elements, however, they are more promising
in the search for high(room)-temperature superconductivity.~\cite{2021SONa,2017MAa,2017MAb,2019SHA,2021ZHE,2017RAH,2021LIA,2019SUN,2019LIAa,2020WEI,2021SHI,2021SONb,2021SEM,2019LIAb,2021JIA,2020CUI,2020CAT,2021SONc,2022CAT,2022VOC,2019SOM}
For example, the DFT calculations by Sun \textit{et al.}~\cite{2019SUN} predicted that magnesium hydride can dissociate the H$_{2}$ molecule into atomic H by lithium doping.
This doping increases
its electron concentration at the Fermi level ($E_\mathrm{F}$), enabling it to reach a $T\rm_{c}$ value of 473 K at 250 GPa.~\cite{2019SUN}
Furthermore, a recent high-pressure experiment has revealed that a small amount of carbon doping in H$_{3}$S can result in
room-temperature superconductivity ($T\rm_{c}$ = 287.7 $\pm$ 1.2 K at 267 $\pm$ 10 GPa),~\cite{2020SNI}
although the anomaly of AC susceptibility in low temperature for the carbonaceous sulfur hydride has been questioned by Hirsch \textit{et al}.~\cite{2021HIR,2022HIR} and more strong experimental data are needed to support the claim of the observation of room-temperature superconductivity.
Therefore, increasing the density of electronic states at $E_\mathrm{F}$ by electrons doping in the binary hydride can affect the $T\rm_{c}$ value; however this modification also can result in a decrease in $T\rm_{c}$ in some cases.~\cite{2018NAK,2019AMS,2021GUA}

\vspace{2mm}
Furthermore, most of the
high-$T\rm_{c}$
superconducting hydrides reported
thus far are concentrated in the Mendeleev table's "lability belt", and many of these
hydrides have unusual structural properties, such as a cage-like structure surrounded by H.~\cite{2020SEM}
Some ternary superconducting hydrides (e.g., ScYH$_{6}$, ScCaH$_{8}$, CaYH$_{12}$,
and so on) can be easily obtained by substituting one of the metal atoms in the binary parent compound.~\cite{2019LIAa,2020WEI,2021SHI,2021SONb,2021SEM}
The crystal structure, pressure range of stability,  unit cell volume, and other
properties of the parent binary superconducting hydrides that
correspond to these ternary superconducting hydrides
resemble each other.~\cite{2020SEM,2020FLO}
For instance, by combining two types of binary hydrides with comparable structural features,
it is possible to produced a more stable ternary hydride with similar structural properties.
To test this hypothesis, we noticed that CeH$_{9}$, YH$_{9}$, and LaH$_{10}$, CeH$_{10}$ have comparable
qualities in a variety of properties, and their structural properties have been proven in high-pressure experiments.~\cite{2019SAL,2021CHEb,2021KON,2019DRO}
The related ternary superconducting hydrides, one of which is (La,Y)H$_{10}$, has been successfully synthesized experimentally, and the $T\rm_{c}$ observed under high pressure also matched the theoretically
predicted $T\rm_{c}$ quite well.~\cite{2021SEM}
During this work, we are aware that a very recent high-pressure experiment study shows that La-Ce-H may have stable $I4/mmm$-(La,Ce)H$_{4}$ and $Im\bar{3}m$-(La,Ce)H$_{6}$ around 100 GPa.~\cite{2022CHEb}

\vspace{2mm}
Therefore, this work will concentrate on the YCeH$_{x}$ and LaCeH$_{x}$ systems, which have not yet been well
investigated, to explore whether there exist stable compounds under high pressure.
By employing the evolutionary algorithm for crystal structure prediction,
we have theoretically discovered thermodynamically stable compounds
of YCeH$_{5}$,YCeH$_{7}$,YCeH$_{8}$,
YCeH$_{18}$, YCeH$_{20}$, LaCeH$_{8}$,
LaCeH$_{18}$, and LaCeH$_{20}$ in pressure range of 100-400 GPa.
In addition, we have investigated the dynamic stability and superconductivity of these hydrogen-rich ternary hydrides within the harmonic approximation.
All of them are dynamically stable with increasing pressure,
More importantly, YCeH$_{18}$ exhibits extremely strong electron-phonon coupling with superconducting transition temperatures of 173 K at 150 GPa.
\section{Method}
The crystal-structure search for YCeH$_{x}$ and LaCeH$_{x}$ ($x$ = 2$\sim$10, 12, 14, 16, 18, and 20) at 100, 200, and 300 GPa was performed using the USPEX (Universal Structure Predictor: Evolutionary Xtallography)~\cite{2006GLA,2013LYA} code.
In the USPEX run, 200 structures with 1-2 formula units were randomly built for the initial generation, and then 100 structures for each of the subsequent generations were produced by 40\% heredity, 40\% random, 10\% mutation, and 10\% soft mutation.
Each structure was optimized through 4 times of relaxation from low to high accuracy levels. The structure relaxation was carried out by the density functional theory calculations within the generalized gradient approximation (GGA) with Perdew-Burke-Ernzerhof (PBE) exchange-correlation functional~\cite{Perdew1996prl} as implemented in the VASP (Vienna \textit{ab initio} simulation package)~\cite{Kresse1996prb,Kresse1999prb} code. The electron-ion interaction was treated using projector-augmented-wave (PAW) potentials.~\cite{Kresse1996prb,Kresse1999prb} The Ce 4f electrons were explicitly considered into the valence electrons.
The cutoff energy for plane waves was set to 600 eV and the smallest allowed spacing between the $k$-points in the irreducible Brillouin zone was set to be 0.2 $\AA^{-1}$.
To study the chemical bonding and electronic structures of some selected stable YCeH$_x$ and LaCeH$_x$, we have carried out the static calculations using the VASP code and then  performed the Crystal Orbital Hamilton Population (COHP) and Crystal Orbital Bond Index (COBI) analyses using the Lobster 4.1.0 package~\cite{2021MUL} together with the pbeVaspFit2015 basis set.
The electronic band structures were also checked by taking into account the spin-orbit coupling (SOC) interaction and the correction of the on-site
Coulomb interaction of Ce 4f orbitals. The DFT+\textit{U} method of Dudarev \textit{et al.}~\cite{1998DBS} was employed in the latter case. The Hubbard \textit{U} parameter for Ce 4f orbitals was determined by the density-functional perturbation theory approach, as implemented in the HP package~\cite{2022TIM}. The obtained \textit{U} for Ce 4f orbitals is 4.2 eV, which is very close to the value (i.e., 4.5 eV) adopted by Wang \textit{et al.}~\cite{2021WAN} for their study on CeH$_9$. The exchange interaction parameter $J$ for Ce 4f orbitals was taken from Ref.~\citenum{2021WAN} and set to 0.5 eV.

\vspace{2mm}
The formation enthalpies of La-Ce-H and Y-Ce-H ternary systems
(herein abbreviated to $A$-Ce-Y with $A$ = La and Y), relative to
the elemental solids Y~\cite{2012CHE}, Ce~\cite{1999VOH}, La~\cite{2022CHE}, and H~\cite{2007PIC},
were calculated at each pressure defined as below:
  \begin{equation}
    \Delta \mathcal{H}=\frac{1}{x+y+z}
    \left[ \mathcal{H}({A}_{x}\mathrm{Ce}_{y}\mathrm{H}_{z})
      - x\mathcal{H}({A})
      - y\mathcal{H}(\mathrm{Ce})
      - z\mathcal{H}(\mathrm{H})\right],
  \end{equation}
  where $\mathcal{H}(A_{x}\mathrm{Ce}_{y}\mathrm{H}_{z})$
  is the enthalpy per formula unit (f.u.) of the A-Ce-H compound;
  $\mathcal{H}(A)$, $\mathcal{H}(\mathrm{Ce})$,
  and $\mathcal{H}(\mathrm{H})$ are the reference enthalpies of $A$, Ce,
  and H single phases, respectively.
  The thermodynamic stability of the predicted phase
  was determined by comparing the formation enthalpies
  with respect to the convex hull energy:
\begin{equation}
E_\mathrm{above\_hull} = \Delta \mathcal{H}  - \Delta \mathcal{H}_{f},
\end{equation}
where $\Delta \mathcal{H}_{f}$ is a convex hull energy
obtained by constraining the minimum value of
the total enthalpies of a linear combination of stable phases~\cite{2007RAK},
which can be computed using the ConvexHull module in scipy.~\cite{2020VIR}
$E_\mathrm{above\_hull}$ is energy above the convex hull.
$E_\mathrm{above\_hull} = 0$ means that
the corresponding ternary phase is stable,
namely, such a phase would not decompose into any combination of elementary,
binary, or other ternary phases.
The Gibbs free energies at finite temperatures were calculated within the quasiharmonic approximation using the PHONOPY code.~\cite{2015TOG}
After incorporating the Gibbs free energies, we can drive the stability for the whole system at high temperatures by computing the convex hull.

\vspace{2mm}
The phonon dispersion and electron-phonon coupling (EPC) of the stable ternary phases were carried out using the Quantum ESPRESSO (QE) code~\cite{Giannozzi2009jcpm} with the PAW method and the GGA-PBE exchange-correlation functional. In particular, the Ce PAW potential was generated with a valence configuration of 4d$^{10}$5s$^{2}$5p$^{6}$4f$^{0.5}$5d$^{1.5}$6s$^{2}$, as provided by the PSLibray of Dal Corso~\cite{2014DalCorso}.
The cutoff energy for plane-wave basis sets in the QE calculations was set to 180 Ry.
The $q$-point mesh (and integral $k$-points mesh) in the first Brillouin-zone for the EPC calculations was set as follows: $5\times5\times2$ ($20\times20\times8$) for $A$CeH$_{8}$, $5\times5\times3$ ($20\times20\times12$) for $A$CeH$_{18}$ and $4\times4\times4$ ($20\times20\times20$) for $A$CeH$_{20}$ ($A$ = La and Y).
The the Allen-Dynes-modified McMillan formula (AD) and the Eliashberg function derived from the EPC calculation were used to predict the superconducting critical temperature as follows~\cite{1968MCM,1975ALL}:
\begin{equation}
\label{eq:AD}
T_\mathrm{c} = \frac{\omega_{\rm{log}} f_{1} f_{2}}{1.2} \rm{exp}\left(\frac{-1.04 (1 + \lambda)}{\lambda (1 - 0.62\mu^{\ast}) - \mu^{\ast}}\right),
\end{equation}
with
\begin{eqnarray}
\label{eq:f1f2}
 \nonumber 
  f_{1}f_{2} &=& \sqrt[3]{1 + \left[\frac{\lambda}{2.46 (1 + 3.8 \mu^{\ast}))}\right]^{\frac{3}{2}}}  \\
    & & \times  \left[1 - \frac{\lambda^{2}(1 - \omega_{2}/\omega_{\rm{log}})}{\lambda^{2} + 3.312(1 + 6.3\mu^{\ast})^{2}}\right].
\end{eqnarray}
When $f_{1}f_{2}$ = 1, Eq.~\eqref{eq:AD} is restored to the original McMillan formula (McM).
The $\mu^{\ast}$ is the Coulomb pseudopotential, which is defined as below~\cite{1975ALL}
\begin{equation}
\mu^{\ast} = \frac{\mu}{1 + \mu \rm{log}(\omega_{el}/\omega_{ph})},
\end{equation}
where $\mu$ is instantaneous repulsion, the $ \rm{log}(\omega_{el}/\omega_{ph})$ is a ratio of propagation time. For most metal compounds, the value of $\mu^{\ast}$ is roughly in the range of 0.1 to 0.16 by incorporating previous experimental and calculated data.
The widely accepted value of 0.1 for $\mu^{\ast}$ was used herein.
The electron-phonon coupling constant ($\lambda$), logarithmic average phonon frequency ($\omega_{\rm{log}}$), and mean square frequency ($\omega_2$) are defined as below
\begin{equation}
\label{eq:epc}
\lambda = 2 \int{\frac{\alpha^{2}F(\omega)}{\omega}}d{\omega},
\end{equation}
\begin{equation}
\label{eq:wlog}
\omega_{\rm{log}} = {\rm{exp}}\left[ \frac{2}{\lambda}\int{\frac{d\omega}{\omega}\alpha^{2}F(\omega){\rm{log}}\omega}      \right ],
\end{equation}
and
\begin{equation}
\label{eq:omega2}
\omega_{2}  = \sqrt{\frac{1}{\lambda} \int{\left[\frac{2\alpha^{2}F(\omega)}{\omega} \right] \omega^{2}d\omega }},
\end{equation}
respectively.

\section{Results and discussion}
\label{sec.results}
\subsection{Phase stability and crystal structural of YCeH$_{x}$ and LaCeH$_{x}$ }
Because predicting the ternary phase diagram in its entirety is computationally expensive,
and our main target is to determine the stability and superconductivity of the hydrogen-rich
phase, this research focuses on stable compounds and crystal structures at 100 GPa, 200 GPa,
and 300 GPa along a line chosen by YCeH$_{x}$ and LaCeH$_{x}$ ($x$ = 2-8, 10, 12, 14, 16, 18, and 20) for the fixed composition search.
This approach is currently seen to be effective for alkaline earths and rare earth metal ternary hydrides.~\cite{2019LIAa,2021SHAa,2020CAT}
For the stability analysis of YCeH$_x$ and LaCeH$_x$, the corresponding Y-H, La-H, and Ce-H binary hydrides have been extensively studied.~\cite{2019DRO,2021CHEa,2021KON}
Theoretical calculations show that there is a hydrogen-rich structure $M$H$_{9}$ ($M$ = Y, La, and Ce) that can be stabilized below
100 GPa in the three binary systems. In addition, YH$_{10}$ is capable of exhibiting room temperature superconductivity.~\cite{2017PEN}
Therefore, this study mainly used these stable binary hydrides as well as single elements as a reference for thermodynamic stability assessment.

\begin{figure*}[htbp]
  \begin{center}
    \includegraphics[width=\linewidth]{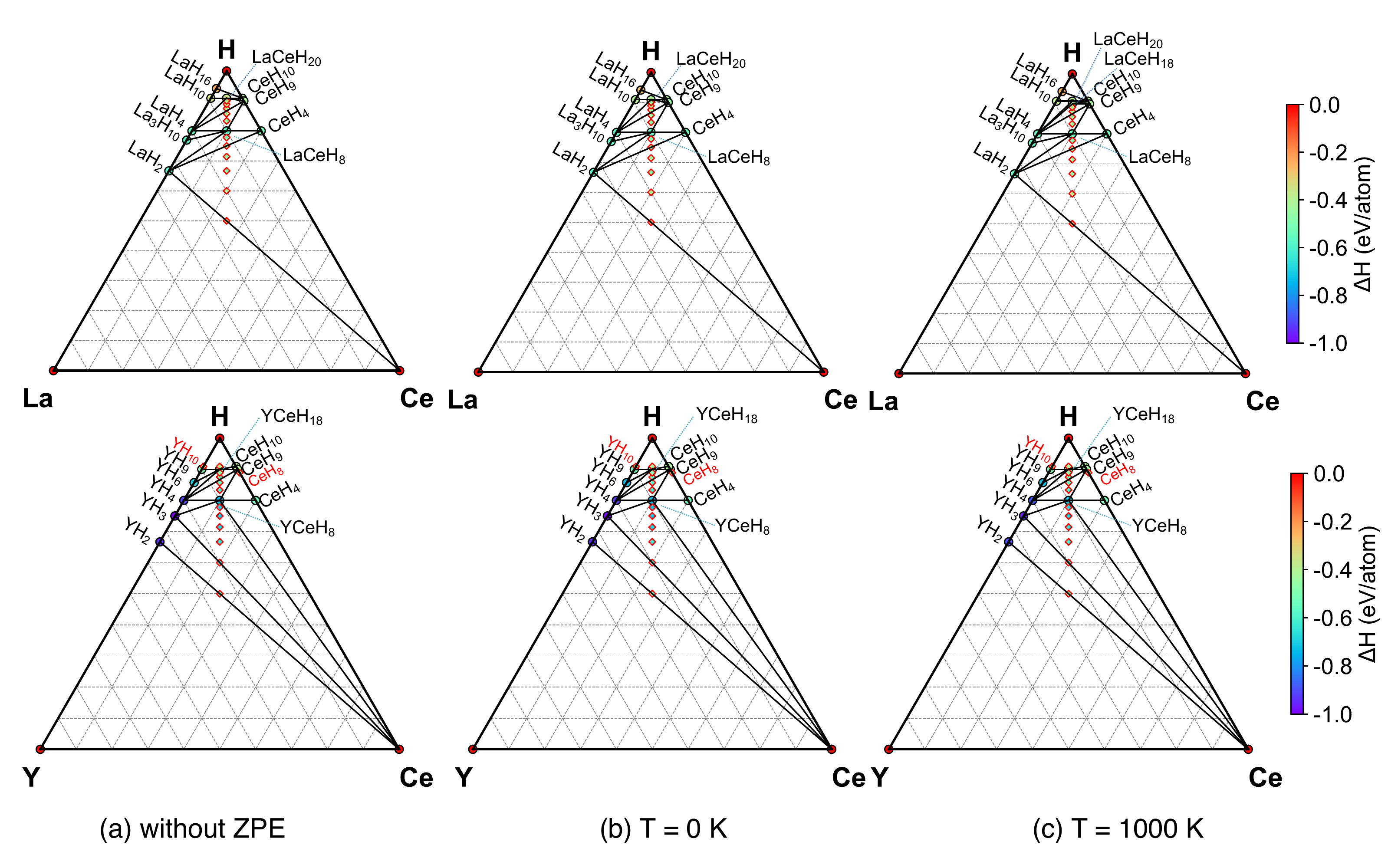}
    \caption{Ternary convex hulls of YCeH$_{x}$ and LaCeH$_{x}$ at 200 GPa. The stable and metastable phases are highlighted by the symbols of circles and red diamond, respectively.  The right panels are the convex hull at 0 K (with the correction of zero-point energy) and 1000 K. }
    \label{fig1.LaCeH_convex_hull}
  \end{center}
\end{figure*}

\begin{figure}[htbp]
  \begin{center}
    \includegraphics[width=\linewidth]{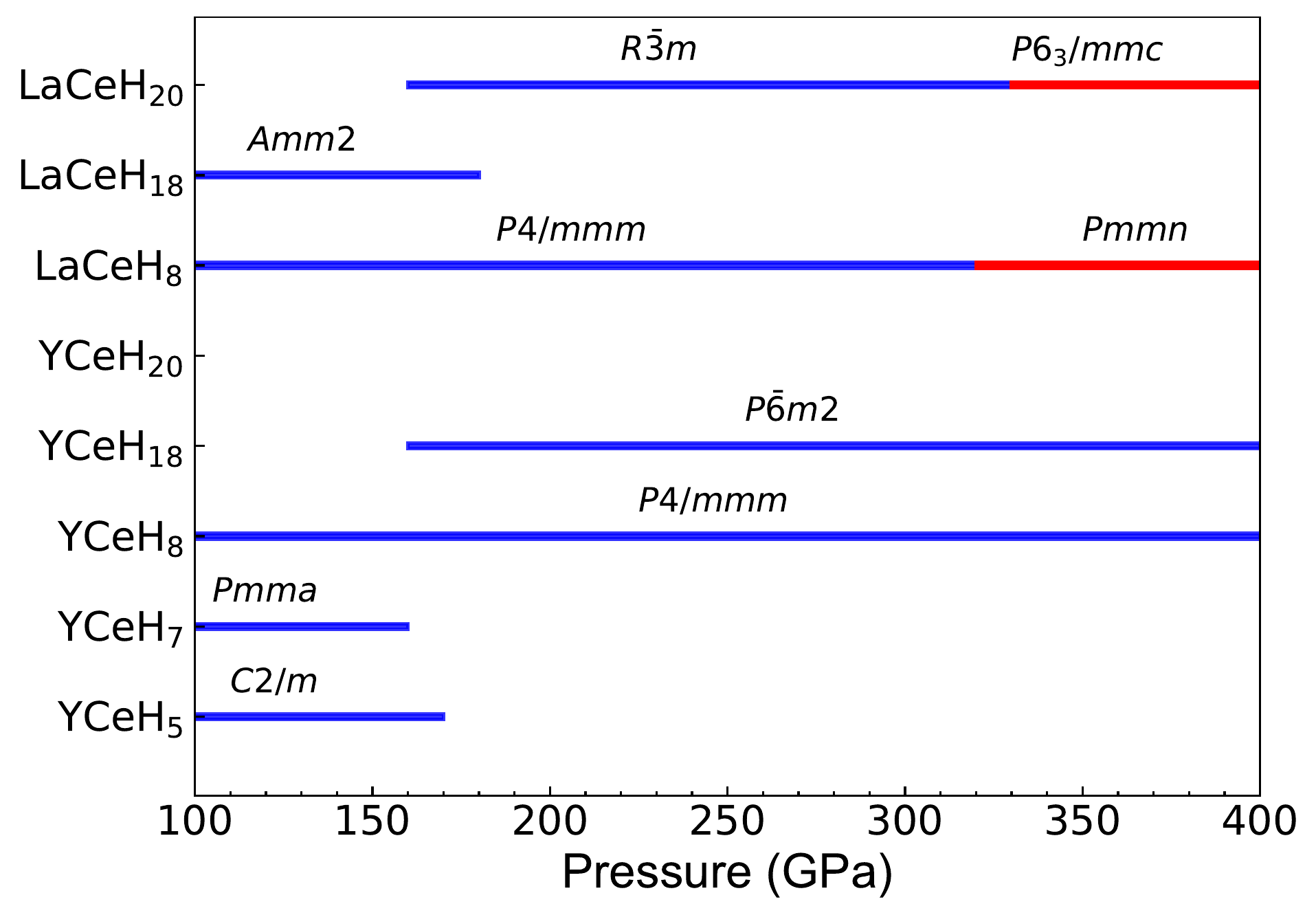}
    \caption{Phase diagram of YCeH$_{x}$ and LaCeH$_{x}$ in the pressure range from 100 to 400 GPa.}
    \label{fig2.phase_diagram}
  \end{center}
\end{figure}

The constructed convex hulls of YCeH$_x$ and LaCeH$_x$ at 200 GPa are presented in Fig.~\ref{fig1.LaCeH_convex_hull}. Their stable and metastable phases are highlighted by the symbols of circles and red diamond in Fig.~\ref{fig1.LaCeH_convex_hull}, respectively. The more detailed results for the stabilites of YCeH$_x$ and LaCeH$_x$ at difference pressures are presented in Figs. S1-S6 in the Supplementary Material (SM).
YCeH$_{5}$, YCeH$_{7}$, YCeH$_{8}$, LaCeH$_{8}$, and LaCeH$_{18}$ are found to be stabilized at 100 GPa. When the pressure increases up to 200 GPa, YCeH$_5$, YCeH$_7$, and LaCeH$_{18}$ vanish from the ternary convex, while LaCeH$_{20}$ and YCeH$_{18}$ appear in the stable phases of convex hull.
In the pressure range of 100-300 GPa, YCeH$_{20}$ fails to achieve stability. But further increasing the pressure above 300 GPa, $R\bar{3}m$-YCeH$_{20}$ becomes stable, and its stable pressure interval is very close to that of YH$_{10}$.~\cite{2017PEN}
In addition, in the pressure interval of 300-400 GPa, LaCeH$_{8}$, LaCeH$_{20}$, and YCeH$_{20}$ undergo a phase transition to the more stable $Pmmn$, $P6_{3}/mmc$, and $P4/mmm$ phases, respectively.
The ranges of the stabilization pressures of these new predicted phases have been summarized in Fig.~\ref{fig2.phase_diagram}.
To further explore the resistance of these phases to decomposition at high temperatures, we calculated the ternary convex hull at a finite temperature at 200 GPa.
We note that the stable phase (YCeH$_{8}$, YCeH$_{18}$, LaCeH$_{8}$, and LaCeH$_{20}$) at this point is also stable at high temperatures, and its specific energy values have been placed in the SM.
Therefore, the above-mentioned can be synthesized in the same way as (La,Y)H$_{x}$~\cite{2021SEM} by heating LaCe and YCe alloys with NH$_{3}$BH$_{3}$ at high pressure.

\begin{figure}[htbp]
  \begin{center}
    \includegraphics[width=\linewidth]{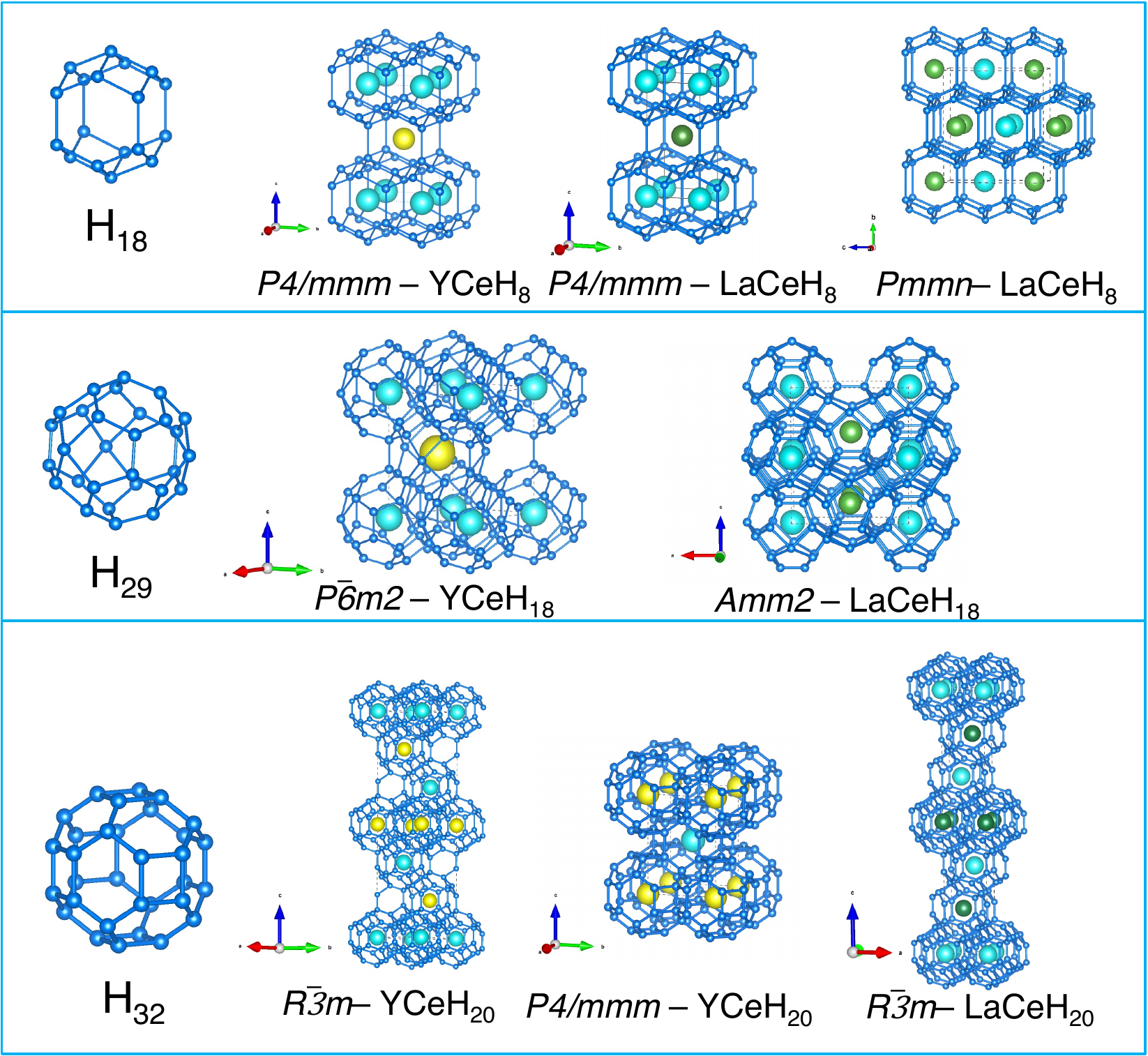}
    \caption{The clathrate structures of stable YCeH$_{x}$ and LaCeH$_{x}$.  }
    \label{fig3.strcuture}
  \end{center}
\end{figure}

It is interesting to note that most of the above predicted stable phases, except YCeH$_{5}$ and YCeH$_{7}$, are composed of clathrate structures enclosed by H. For instance, the clathrate structures of stable YCeH$_8$, LaCeH$_8$, YCeH$_{18}$, LaCeH$_{18}$, and YCeH$_{20}$ are shown in Fig.~\ref{fig3.strcuture}.
Since the hydrogen-rich compounds are more likely to achieve high-$T\rm_{c}$
superconductivity~\cite{2020FLO}, the hydrogen-less ones such as YCeH$_{5}$ and YCeH$_{7}$ will be discussed elsewhere.
The stable hydrides of YCeH$_x$ and LaCeH$_x$ with clathrate structures can be roughly divided into three groups.

\vspace{2mm}
(i) The first one consists of H$_{18}$ cages, as found in $P4/mmm$-YCeH$_{8}$, $P4/mmm$-LaCeH$_{8}$, and $Pmmn$-LaCeH$_{8}$, which are derived from their corresponding parent compounds ($I4/mmm$-YH$_{4}$, $I4/mmm$-CeH$_{4}$, and $I4/mmm$-LaH$_{4}$)~\cite{2017PEN} with the same cage structure. If half of Ce atoms at the site $2a$ of the $I4/mmm$-CeH$_{4}$ phase are orderly replaced by Y or La atoms, the $P4/mmm$ ternary phase would be formed.~\cite{2017PEN} The relative enthalpies of these two structures of LaCeH$_{4}$ and YCeH$_{4}$ are shown in Figs. S1 and S4.
We find that the most stable candidate structures prefer the H$_{18}$ cages.
In addition, LaCeH$_{8}$ would be transformed into the $Pmmn$ phase when the pressure is above 300 GPa.
(ii) The second group consists of H$_{29}$ cages, as found in the $P\bar{6}m2$-YCeH$_{18}$ and $Amm2$-LaCeH$_{18}$ phases. The $P\bar{6}m2$-YCeH$_{18}$ phase can be derived by replacing half of Ce atoms at the site 2$d$  of $P6_{3}/mmc$-CeH$_{9}$ with Y atoms.
As for the parent compounds of $P\bar{6}m2$-YCeH$_{18}$ and $Amm2$-LaCeH$_{18}$, the previous study by Peng \textit{et al}.~\cite{2017PEN} has shown that only stable $P6_{3}/mmc$-YH$_{9}$ and $P6_{3}/mmc$-CeH$_{9}$ phases exist in the pressure range of 100-400 GPa, however LaH$_{9}$ does not.
For the $Amm2$-LaCeH$_{18}$ phase, only its internal energy plays a superior role to hinder its possible decomposition pathway of 1/6 LaH$_{4}$ + 1/6 LaH$_{10}$ + CeH$_{9}$, and thus it cannot be stabilized at higher pressures.~\cite{2017LIU,2017PEN}
(iii) The third group consists of the H$_{32}$ cages, as found in YCeH$_{20}$ and LaCeH$_{20}$. The $R\bar{3}m$-YCeH$_{20}$ and $R\bar{3}m$-LaCeH$_{20}$ phases exhibit the same spatial structure with LaYH$_{20}$,~\cite{2021SEM} and all of them are the supercell structures of their parent phases $Fm\bar{3}m$-$A$H$_{10}$. More specifically speaking, the equivalent primitive unit cell of $R\bar{3}m$-YCeH$_{20}$ can be obtained by replacing half of Ce atoms with Y atoms in the $1\times1\times2$ extension of the primitive unit cell of $Fm\bar{3}m$-CeH$_{10}$.
LaCeH$_{20}$ undergoes a phase transition from $R\bar{3}m$ to $P6_{3}/mmc$.
For YCeH$_{20}$, it also undergoes a phase transition at 370 GPa and the new high-pressure phase can be derived by replacing half of Ce atoms in the site $4b$ of the parent phase $Fm\bar{3}m$-CeH$_{10}$ with Y atoms. Considering the chemical similarity of elements, one may choose the stable phases of binary hydrides as the starting structures to accelerate the search for multi-component hydrides.

To further check the lattice dynamic stability of the above predicted YCeH$_x$ and LaCeH$_x$ phases, we have employed the Phonopy~\cite{2015TOG} code to initially calculate their phonon band structures, which are presented in Fig.~S7 in the SM.
Among them, the $Amm2$-LaCeH$_{18}$  phases cannot maintain dynamic stability in the pressure range of their thermodynamic stability.
To make the $R\bar{3}m$-LaCeH$_{20}$ phase dynamically stable, the pressure needs to be above 200 GPa.
In addition, the high pressure applied in the current mainstream experimental studies of metal hydrides falls into a range of 100-300 GPa. As for the phases stabilized above 300 GPa, both $R\bar{3}m$-YCeH$_{20}$ and $R\bar{3}m$-LaCeH$_{20}$ have a certain degree of similarity.
Therefore, we pay much attention to the $P4/mmm$-YCeH$_{8}$, $P\bar{6}m2$-YCeH$_{18}$, $R\bar{3}m$-YCeH$_{20}$, $P4/mmm$-LaCeH$_{8}$, and $R\bar{3}m$-LaCeH$_{20}$ phases for their electronic properties and superconductivity.

\subsection{Electronic Properties and Superconductivity}
\vspace{2mm}
The electronic band structure, the electronic density of states (eDOS), of $P4/mmm$-YCeH$_{8}$, $P\bar{6}m2$-YCeH$_{18}$, $R\bar{3}m$-YCeH$_{20}$, $P4/mmm$-LaCeH$_{8}$, and $R\bar{3}m$-LaCeH$_{20}$ at high pressures are presented in Figs.~S8-S10 in the SM, respectively. All these compounds in the pressure range of their thermodynamical stabilities possess the metallic features in their energy band structures. For $P4/mmm$-YCeH$_{8}$ and $P4/mmm$-LaCeH$_{8}$, there are very steep conduction bands along the $Z\rightarrow\Gamma$ direction crossing the Fermi level ($E_\mathrm{F}$), resulting in electron pockets near the $\Gamma$ point. For $R\bar{3}m$-YCeH$_{20}$ and $R\bar{3}m$-LaCeH$_{20}$, a band inversion can be found around the $\Gamma$ point and near the $E_\mathrm{F}$, which is mainly attributed to the H 1s and Ce 4f orbitals, as seen from the atom- and orbital-weighted band structures in Figs.~S8 and~S9 in the SM, respectively. The eDOS at $E_\mathrm{F}$ ($N_{E_\mathrm{F}}$) for $P4/mmm$-YCeH$_{8}$, $P\bar{6}m2$-YCeH$_{18}$, $R\bar{3}m$-YCeH$_{20}$, $P4/mmm$-LaCeH$_{8}$, and $R\bar{3}m$-LaCeH$_{20}$ are 1.72, 1.39, 1.10, 1.34, and 1.26 states/eV/f.u., respectively.
As further analyzed by the atom-projected eDOS, the difference in the aforementioned values of $N_{E_\mathrm{F}}$ mainly comes from the different contribution of H and Ce atoms to the $N_{E_\mathrm{F}}$.
The contribution of Ce atoms at $E_\mathrm{F}$ is continuously suppressed as the hydrogen content increases, while the opposite trend is observed for H.
However, the decrease of Ce is higher than the increase of H, leading to a constant decrease of the total DOS with increasing H content.
According to the arguments of Belli \textit{et al.},~\cite{2021BNCE} the superconductivity of hydrides maintains a strong positive correlation with the contribution of H in the total DOS at $E_\mathrm{F}$ ($H_\mathrm{DOS}$). The $H_\mathrm{DOS}$ for these five phases are 9.5\%, 34.4\%, 38.2\%, 11.2\%, and 34.7\%, respectively.
This indicates that the H-rich phase has a higher H-derived DOS at $E_\mathrm{F}$ although the corresponding total DOS is low, which may suggest that the possibility of high-$T\rm_{c}$ in H-rich phases to some extent.~\cite{2021BNCE}

\vspace{2mm}
From the orbital-weighted electronic band structures of $P\bar{6}m2$-YCeH$_{18}$, $R\bar{3}m$-YCeH$_{20}$ and, $R\bar{3}m$-LaCeH$_{20}$, as shown in Fig.~S9, it can be seen that the bands around $E_\mathrm{F}$ are mainly contributed by the hybridization between the Ce 4f and H 1s orbitals. We should point out that they were predicted by the GGA-PBE functional. This raises a question how strong the SOC interaction and the correlated effect would be associated with the Ce 4f orbitals in these compounds. So, we further checked the electronic band structures with the SOC interaction, the correction of the on-site Coulomb interaction of Ce 4f orbitals, and their  combination, which are presented in Fig.~S10 in the SM. We find that both the SOC interaction and the correction of the on-site Coulomb interaction have much weak effect on the bands around $E_\mathrm{F}$, which could be neglected. The superconducting transition temperature is largely affected by the electronic properties at the $E_\mathrm{F}$. Therefore, the computational methodology employed in the present study would be acceptable to give a reasonable prediction on the superconducting properties of Ce-containing ternary hydrides. The similar treatment with ignoring the SOC interaction and the correlated effect has also been employed in several of the previous theoretical studies~\cite{2017PEN,2021WAN,2019LMT} on the electron-phonon coupling of binary CeH$_x$. Unfortunately, the combination of the DFT+\textit{U} method and the density-functional perturbation theory approach in the latest version (v7.1) of QE code does not support the electron-phonon interaction calculations. Because of the above reasons, our superconductivity calculations did not consider the SOC interaction and the strongly-correlated effect.

\begin{figure*}[htbp]
  \begin{center}
    \includegraphics[width=\linewidth]{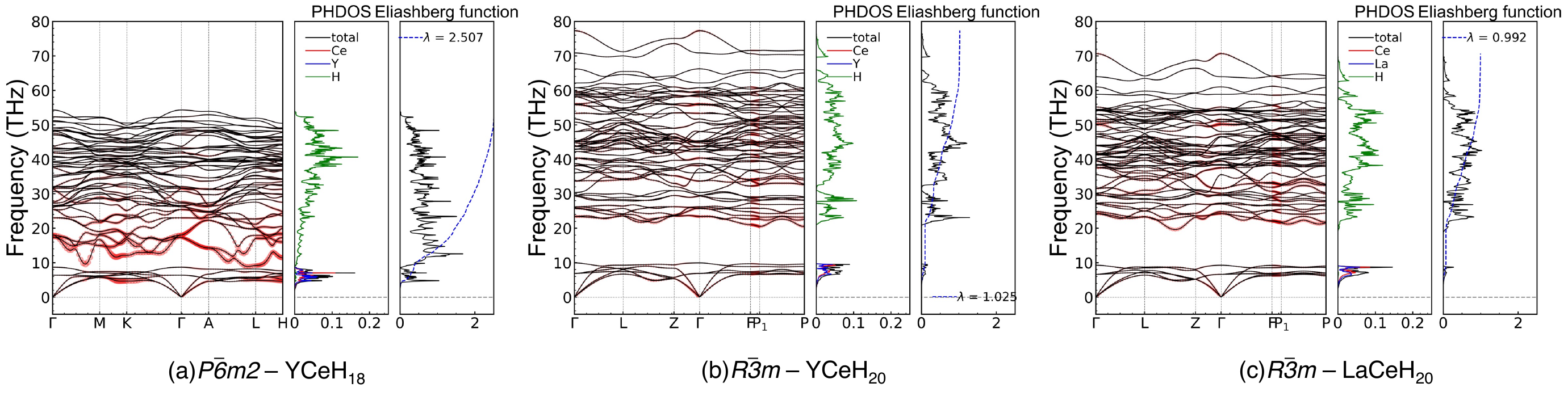}
    \caption{The phonon spectra with a projection of mode-resolved EPC constant $\lambda_{q\nu}$ (red circle), atom-projected phonon density of states (PHDOS), and Eliashberg function of (a) $P\bar{6}m2$-YCeH$_{18}$ at 150 GPa, (b) $R\bar{3}m$-YCeH$_{20}$ at 300 GPa, and (c) $R\bar{3}m$-LaCeH$_{20}$ at 250 GPa. The mode-revolved EPC constant $\lambda_{q\nu}$ is defined as
$\lambda_{q\nu} = \frac{\gamma_{q\nu}}{\pi\hbar N(E_{F}) \omega_{q\nu}^{2}}$, where $\gamma_{q\nu}$ is phonon linewidth, $N(E_{F})$ is the DOS at the Fermi level, and $\omega_{q\nu}$ is the mode-resolve phonon frequency.}
    \label{fig4.EPC}
  \end{center}
\end{figure*}

\vspace{2mm}
The phonon dispersions with the mode-resolved electron-phonon coupling (EPC) constant $\lambda_{q\nu}$, atom-projected phonon density of states (PHDOS), and EPC spectra of $P4/mmm$-YCeH$_{8}$, $P\bar{6}m2$-YCeH$_{18}$, $R\bar{3}m$-YCeH$_{20}$, $P4/mmm$-LaCeH$_{8}$, and $R\bar{3}m$-LaCeH$_{20}$ at high pressures are presented Fig.~\ref{fig4.EPC} and Fig.~S12 in the SM.
For these five phases, the phonons with frequencies above 10 THz are dominated by the H contribution, accounting for more than 70\%. The phonons with frequencies below 10 THz are mainly contributed by La, Ce, and Y atoms.
Because of the difference in the atomic masses of Y, La, and Ce, the low-frequency range in the phonon spectrum of YCeH$_x$ is slightly wider than that of LaCeH$_{x}$.
The EPC constants of YCeH$_8$ at 100 GPa, YCeH$_{18}$ at 150 GPa, YCeH$_{20}$ at 300 GPa, LaCeH$_8$ at 100 GPa, and LaCeH$_{20}$ at 250 GPa are 0.41, 2.51, 1.02, 0.35, and 0.99, respectively.
The EPC constant of YCeH$_{18}$ is so significant compared to the other four phases.
From the mode-resolved EPC constant $\lambda_{q\nu}$, we find that the largest contribution to the total EPC constant of YCeH$_{18}$ comes from the optical phonon branch in the frequency range of 10 to 20 THz, which contributes about 0.64 (i.e., 25.7\% to the total EPC constant).
The rare-earth-atom associated vibrational modes with the frequencies of $<$10 THz for YCeH$_{18}$ also have moderate contribution to the total EPC constant, as compared with the other four phases. In contrast, the contribution of rare-earth atoms to the EPC for YCeH$_{20}$ with higher H content is only 0.09.
When the pressure is increased from 150 GPa to 300 GPa, the phonons dominated by the H-atom-associated vibration in YCeH$_{18}$ keep hardening, which can been seen from Fig.~\ref{fig4.EPC}(a) and Fig.~S11(c). The increase of pressure leads to a decrease of the EPC constant of YCeH$_{18}$, so that the EPC constant of YCeH$_{18}$ at 300 GPa is very close to that of LaCeH$_{20}$ at 300 GPa. The EPC constants of $A$CeH$_{8}$ ($A$ = Y and La) are so small, compared to $A$CeH$_{18}$ and $A$CeH$_{20}$, since both the highest phonon frequencies and the H-derived electronic density of states in the former cases are lower than those in the latter cases. 
These results suggest that once the hydrogen atoms in metal hydrides simultaneously provide sufficient electron density of states at $E_\mathrm{F}$ and sufficient phonon density of states in the high-frequency region, a strong EPC strength would be achieved.

\vspace{2mm}
Based on the Allen-Dynes-modified McMillan formula~\cite{1975ALL} with a widely accepted value of the Coulomb pseudopotential $\mu^{*}$ ($\mu^{*}$ = 0.1 used herein),
the $T_{c}$ values of $P4/mmm$-YCeH$_{8}$ at 100 GPa, $P\bar{6}m2$-YCeH$_{18}$ at 150 GPa, $R\bar{3}m$-YCeH$_{20}$ at 300 GPa, $P4/mmm$-LaCeH$_{8}$ at 100 GPa, and $P4/mmm$-LaCeH$_{8}$ at 250 GPa are predicted to be 4.8, 173.8, 122.1, 1.8, and 116.1 K, respectively.
We note that very recently three independent high-pressure experiments~\cite{2022CHEb,2022BI,2022HUA} on the La-Ce-H system have been reported, in which the measurement pressures by Chen \textit{et al.}~\cite{2022CHEb} and Bi \textit{et al.}~\cite{2022BI} were both less than 130 GPa and the observed $T_\mathrm{c}$ was around 180 K.
One interesting thing is that even though these two research groups~\cite{2022CHEb,2022BI} used different ratios of the starting materials for synthesis, the final measurement results were very close, suggesting that both of them synthesized probably the same high-temperature superconducting phase.
Our calculation results show that at this pressure (130 GPa) the thermodynamically stable phase is LaCeH$_{18}$, which is not explored further here because the stabilization of this phase may involve anharmonic approximation effect not the focus of this work.
The measurement pressure on La-Ce-H by Huang \textit{et al.}~\cite{2022HUA} was in the range of 140-160 GPa. As mentioned above, LaCeH$_{20}$ can be stabilized in this pressure range.
A high-pressure experiment study~\cite{2022CHEc} on the Y-Ce-H system has also been reported very recently, in which the observed $T_{c}$ values are between 97-140 K. The predicted $T_\mathrm{c}$ value of YCeH$_{18}$ by either the McM or AD formula shows a good agreement with this experiment data.
To a certain extent, these experiment results show strong support to the validity of our theoretical prediction.
In short, the high-pressure phases of LaCeH$_{18}$, LaCeH$_{20}$, and YCeH$_{18}$ discovered computationally in this work have all been confirmed in the very recent high-pressure experiments, among which LaCeH$_{20}$ and YCeH$_{18}$ are theoretically predicted for the first time.

\subsection{Discussion}
\begin{figure}[htbp]
  \begin{center}
    \includegraphics[width=\linewidth]{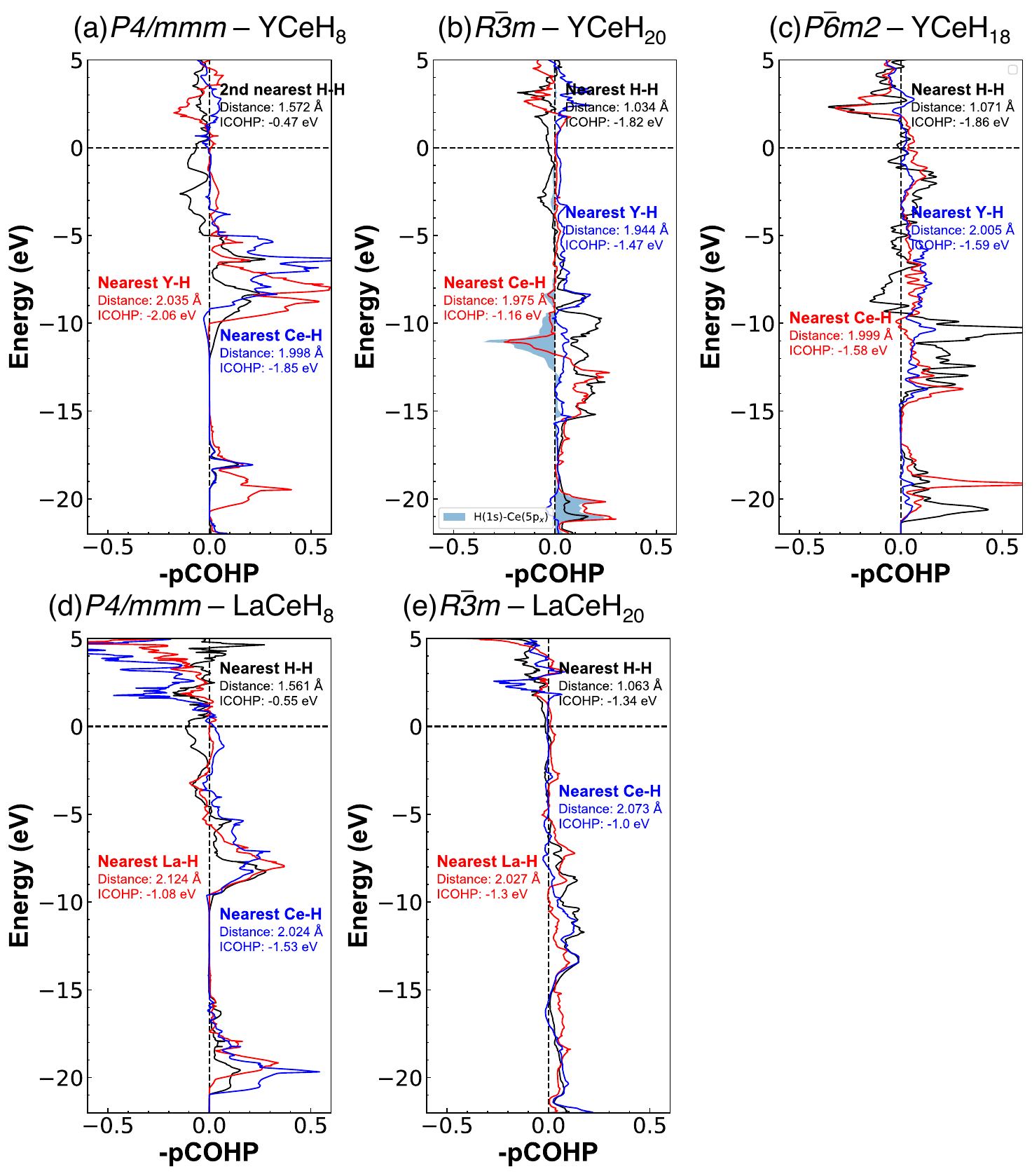}
    \caption{The Crystal Orbital Hamilton Population (COHP) of some representative atom pairs (H-H, Y-H, La-H, and Ce-H) in $P4/mmm$-YCeH8, $P\bar{6}m2$-YCeH$_{18}$, $R\bar{3}m$-YCeH$_{20}$, $P4/mmm$-LaCeH$_{8}$, and $R\bar{3}$m-LaCeH$_{20}$ at the pressure of 100, 150, 300, 100, and 250 GPa, respectively.}
    \label{fig5.cohp}
  \end{center}
\end{figure}

As presented above, $A$CeH$_{8}$, YCeH$_{18}$ and $A$CeH$_{20}$ ($A$ = La and Y) have the space groups of $P4/mmm$, $P\bar{6}m2$ and $R\bar{3}m$, respectively, and the rare earth atoms are encased in the H$_{18}$, H$_{29}$ and H$_{32}$ cages, respectively.
Herein we performed the COHP and integrated COHP (ICOHP) analyses to further investigate the chemical bonding in these cage structures.
The critical bond lengths of some representative H-H, Ce-H, and $A$-H ($A$ - La, Y) atom pairs and their corresponding ICOHP values that can reflect the bonding strength are shown in Figure \ref{fig5.cohp}.
According to the convention of COHP analysis, the positive and negative values of $-$COHP indicate the bonding and anti-bonding characteristics, respectively.
As seen from Fig.~\ref{fig3.strcuture}, the H$_{18}$ cage in YCeH$_{8}$ can be viewed as consisting of eight quadrangles and four hexagons. Accordingly, the H-H bond in the H$_{18}$ cages can be roughly classified into three groups: i) H-H bonds shared by two neighboring quadrangles, ii) H-H bonds shared by two neighboring hexagons, and iii) H-H bonds shared by a quadrangle and a hexagon.

\vspace{2mm}
The H-H bond lengths in the first and second groups are denoted as D1 and D2, respectively, as indicated in Figs. S14(a) and S14(d) in the SM. These two distinct bond lengths D1 and D2 in YCeH$_8$ and LaCeH$_8$ at 100 GPa can be sorted in the following descending order: D2(LaCeH$_8$) $>$ D1(YCeH$_8$) $\gtrsim$ D1(LaCeH$_8$) $>$ D2(YCeH$_8$). Although the H-H bonds with D1 in YCeH$_8$ and LaCeH$_8$ have the very close bond lengths, the projected COHPs for them shown in Figs.~\ref{fig5.cohp}(a) and \ref{fig5.cohp}(c) indicate that this H-H bond in LaCeH$_8$ exhibits a stronger bonding interaction than in YCeH$_8$ because of the less anti-bonding states appearing just below $E_\mathrm{F}$ and a slightly greater absolute value of the corresponding ICOHP in LaCeH$_8$. However, compared to the $A$-H and Ce-H bonds from their corresponding ICOHP values, the interaction of this H-H bond would be significantly weaker. For YCeH$_{18}$ and $A$CeH$_{20}$ with higher H content, the H-H bonds are significantly shrunken because of a higher pressure required to stabilize YCeH$_{18}$ and $A$CeH$_{20}$, the ICOHP values of the H-H bonds in YCeH$_{18}$ and ACeH$_{20}$ are almost three times than those in $A$CeH$_8$, and much less (and even no) anti-bonding states appear below $E_\mathrm{F}$ for the H-H bonds in YCeH$_{20}$ (and LaCeH$_{20}$). From the H$_{18}$ cages in $A$CeH$_8$ to the H$_{29}$ cages in YCeH$_{18}$ and H$_{32}$ cages in $A$CeH$_{20}$, the cage size increases and the number of quadrangles (hexagons) decreases (increases) due to incorporating more H atoms. These results suggest that the H-H bonds within anti-bonding states in $A$CeH$_8$ could be much stabilized in YCeH$_{18}$ and $A$CeH$_{20}$ because they seem ready to be hybridized with the incorporated additional H atoms. In addition, YCeH$_{20}$ has some remarkable anti-bonding states appearing around -11 eV for the Ce-H atom pair, as seen from Fig.~\ref{fig5.cohp}(c), which mainly come from the hybridization between the H-1s and Ce-5p$_x$ orbitals, suggesting that the semi-core states (such as 5p orbitals) of Ce in YCeH$_{20}$ at the high pressure may contribute to the Ce-H bonding.

\vspace{2mm}
The crystal orbital bond index (COBI) can be used to quantify the strength of covalent bonds in covalently bonded solid materials.~\cite{2021MUL} The integrated-COBI (abbreviated to ICOBI) values for the H-H bonds in these caged structures of $A$CeH$_8$, YCeH$_{18}$ and $A$CeH$_{20}$ are less than 0.15, indicating very weak bonding interactions between H-H. The overall trend is that the ICOBI of H-H bonds increases with the increase of pressure, but the change is less than 0.02, suggesting that the effect of pressure on the H-H bonding is small. The similar trend is also observed from the analysis of ICOHP. One interesting thing is that the bonding strength of the H-H bonds in $R\bar{3}m$-YCeH$_{20}$ in the pressure range of 300-400 GPa is slightly stronger than the one in $R\bar{3}m$-LaCeH$_{20}$ according to both the electron localization function (ELF) and ICOHP analyses, although the corresponding ICOBI values in these two structures are quite close to each other and also exhibit a very weak dependence on the pressure.

\vspace{2mm}
The five superconducting compounds reported above, which are both thermodynamically and dynamically stable, are mainly of the $A$CeH$_{8}$, $A$CeH$_{18}$ and $A$CeH$_{20}$ types.
The previously reported ScCaH$_{8}$~\cite{2021SHI} and YMgH$_{8}$~\cite{2021SONb} also have the same structure as $A$CeH$_{8}$. For the $A$CeH$_{8}$ type, there are three distinct types of H-H bonds, as mentioned above, in which the D1 decreases as the pressure increases, while the D2 exhibits a weaker dependence on the pressure.
In additional, the bonding strength of the H-H bonds can also be described as the ELF.

\vspace{2mm}
Recently, Belli \textit{et al.}~\cite{2021BNCE} have proposed a strong correlation between the $T_\mathrm{c}$ value of superconducting hydride and the weak covalent H-H bonds. They also proposed a possible networking value $\phi$, which is defined as the highest value of the ELF that creates an isosurface spanning through the whole crystal in all three Cartesian directions, to describe the nature of the H-H bond in hydrides.
As noted by Belli \textit{et al.},~\cite{2021BNCE} the precise acquisition of the $\phi$ value is not uniquely defined. From the radial distribution function of H atoms in our studied YCeH$_x$ and LaCeH$_x$ phases as a function of pressure, as presented in Fig.~S14 in the SM, we note that within the radial range of $3$ \AA~there is a main peak representing the most important distribution of H-H bond lengths. In this way, we obtain the $\phi$ value by taking the average of all ELF values at the valley points of the lines plots of ELF along different H-H bonds that correspond to such a main peak.
And according to the model proposed by Belli \textit{et al.},~\cite{2021BNCE} the $T_\mathrm{c}$ of metal hydride can be predicted by the following equation with an accuracy in the range of 60 K:
\begin{equation}
\label{eq:Tc_model}
T_\mathrm{c} = 750  \phi H_{f} \sqrt[1/3]{H_\mathrm{DOS}} - 85,
\end{equation}
where $H_{f}$ is the hydrogen fraction and $H_\mathrm{DOS}$ is the hydrogen fraction of the total DOS at $E_\mathrm{F}$.
As presented in Table~\ref{table.Pdep}, the network values $\phi$ of our studied phases are all less than 0.5 and they are very close. Consequently, the $T_\mathrm{c}$ values predicted by Eq.~\eqref{eq:Tc_model} are more significantly affected by the contribution of $H_\mathrm{DOS}$. We also noted that the $T_\mathrm{c}$ values predicted by Eq.~\eqref{eq:Tc_model} are still somewhat deviation from those predicted by the AD formula, i.e., Eq.~\eqref{eq:AD}.
Through the examination on YCeH$_{18}$ at 150 and 300 GPa, it can be seen that when the pressure increases, the H-H distance decreases, the overall network $\phi$ value exhibits an increasing trend, and the contribution of H to the density of states at $E_\mathrm{F}$ also increases. And thus the Eq.~\eqref{eq:Tc_model} based on ELF predicts that the $T_\mathrm{c}$ for YCeH$_{18}$ increases with the increase of pressure, which is opposite to the trend in the prediction of the AD formula.
This indicates that one possible limitation in the use of Eq.~\eqref{eq:Tc_model} to predict $T_\mathrm{c}$ may be the pressure-dependence of $T_\mathrm{c}$.
From the predicted results of Peng \textit{et al.}~\cite{2017PEN} for all binary rare-earth metal hydrides, a trend can be seen that the elements containing more f electrons exhibit a suppressing effect on the $T_\mathrm{c}$ value. However, the dataset to build Eq.~\eqref{eq:Tc_model} in the work of Belli \textit{et al.}~\cite{2021BNCE} contains too few the f-electron-containing cases (i.e., only CeH$_9$ and PrH$_9$ included therein). 
Therefore, the $T_\mathrm{c}$ does exhibit a certain relationship with the $\phi$ and $H_\mathrm{DOS}$ values, however, the generalization of Eq.~\eqref{eq:Tc_model} to the pressure-dependence and the f-electron-containing systems needs a further improvement.

\section{Conclusion}
\label{sec.conc}
In summary, we employed the evolutionary algorithm and first-principles calculations to study the ternary phase diagrams of YCeH$_{x}$ and LaCeH$_{x}$ systems under the pressure range of 100-400 GPa.
The clathrate structure surrounded by H$_{18}$, H$_{29}$, and H$_{32}$ can maintain thermodynamic stability in the searched pressure range.
These cage structures correspond to YCeH$_{8}$, YCeH$_{18}$, YCeH$_{20}$, LaCeH$_{8}$, LaCeH$_{18}$, LaCeH$_{20}$, respectively.
However, under the harmonic approximation,  LaCeH$_{18}$ cannot achieve dynamic stability.
We further studied the superconductivity of YCeH$_{8}$, YCeH$_{18}$,YCeH$_{20}$, LaCeH$_{8}$, and LaCeH$_{20}$. The results show that all four can achieve high temperature superconductivity, and $P\bar{6}m2$-YCeH$_{18}$ can have the highest 173 K at 150GPa.
In addition, in terms of the structure of these phases, $P4/mmm$-YCeH$_{8}$, $P4/mmm$-LaCeH$_{8}$, $P\bar{6}m2$-YCeH$_{18}$, $P4/mmm$-YCeH$_{20}$ can all be obtained by element replacement in binary superconducting hydrides.
In other words, the search for new superconducting hydrides can start from the combination of binary superconducting hydrides. When two kinds of binary superconducting hydrides with similar properties are combined together, it is possible to form a novel stable three ternary hydride.

\section*{Acknowledgments}
The computations in this work have been performed
using the facilities of
Research Center for Advanced Computing
Infrastructure (RCACI) at JAIST.
P.S. is grateful for financial support from Grant-in-Aid for
JSPS Research Fellow (JSPS KAKENHI Grant No. 22J10527).
K.H. is grateful for financial support from
the HPCI System Research Project (Project ID: hp190169) and
MEXT-KAKENHI (JP16H06439, JP17K17762, JP19K05029, and JP19H05169).
R.M. is grateful for financial supports from
MEXT-KAKENHI (19H04692 and 16KK0097),
FLAGSHIP2020 (project nos. hp1
90169 and hp190167 at K-computer),
Toyota Motor Corporation, I-O DATA Foundation,
the Air Force Office of Scientific Research
(AFOSR-AOARD/FA2386-17-1-4049;FA2386-19-1-4015),
and JSPS Bilateral Joint Projects (with India DST).

\bibliographystyle{apsrev4-1}
\bibliography{references}

\begin{thebibliography}{94}%
\makeatletter
\providecommand \@ifxundefined [1]{%
 \@ifx{#1\undefined}
}%
\providecommand \@ifnum [1]{%
 \ifnum #1\expandafter \@firstoftwo
 \else \expandafter \@secondoftwo
 \fi
}%
\providecommand \@ifx [1]{%
 \ifx #1\expandafter \@firstoftwo
 \else \expandafter \@secondoftwo
 \fi
}%
\providecommand \natexlab [1]{#1}%
\providecommand \enquote  [1]{``#1''}%
\providecommand \bibnamefont  [1]{#1}%
\providecommand \bibfnamefont [1]{#1}%
\providecommand \citenamefont [1]{#1}%
\providecommand \href@noop [0]{\@secondoftwo}%
\providecommand \href [0]{\begingroup \@sanitize@url \@href}%
\providecommand \@href[1]{\@@startlink{#1}\@@href}%
\providecommand \@@href[1]{\endgroup#1\@@endlink}%
\providecommand \@sanitize@url [0]{\catcode `\\12\catcode `\$12\catcode
  `\&12\catcode `\#12\catcode `\^12\catcode `\_12\catcode `\%12\relax}%
\providecommand \@@startlink[1]{}%
\providecommand \@@endlink[0]{}%
\providecommand \url  [0]{\begingroup\@sanitize@url \@url }%
\providecommand \@url [1]{\endgroup\@href {#1}{\urlprefix }}%
\providecommand \urlprefix  [0]{URL }%
\providecommand \Eprint [0]{\href }%
\providecommand \doibase [0]{http://dx.doi.org/}%
\providecommand \selectlanguage [0]{\@gobble}%
\providecommand \bibinfo  [0]{\@secondoftwo}%
\providecommand \bibfield  [0]{\@secondoftwo}%
\providecommand \translation [1]{[#1]}%
\providecommand \BibitemOpen [0]{}%
\providecommand \bibitemStop [0]{}%
\providecommand \bibitemNoStop [0]{.\EOS\space}%
\providecommand \EOS [0]{\spacefactor3000\relax}%
\providecommand \BibitemShut  [1]{\csname bibitem#1\endcsname}%
\let\auto@bib@innerbib\@empty
\bibitem [{\citenamefont {Tinkham}(2004)}]{2004MIC}%
  \BibitemOpen
  \bibfield  {author} {\bibinfo {author} {\bibfnamefont {M.}~\bibnamefont
  {Tinkham}},\ }\href@noop {} {\emph {\bibinfo {title} {Introduction to
  superconductivity}}}\ (\bibinfo  {publisher} {Courier Corporation},\ \bibinfo
  {year} {2004})\BibitemShut {NoStop}%
\bibitem [{\citenamefont {Gennes}(2018)}]{2018GEN}%
  \BibitemOpen
  \bibfield  {author} {\bibinfo {author} {\bibfnamefont {P.~G.~D.}\
  \bibnamefont {Gennes}},\ }\href {\doibase 10.1201/9780429497032} {\emph
  {\bibinfo {title} {Superconductivity of Metals and Alloys}}}\ (\bibinfo
  {publisher} {{CRC} Press},\ \bibinfo {year} {2018})\BibitemShut {NoStop}%
\bibitem [{\citenamefont {Linder}\ and\ \citenamefont
  {Robinson}(2015)}]{2015LIN}%
  \BibitemOpen
  \bibfield  {author} {\bibinfo {author} {\bibfnamefont {J.}~\bibnamefont
  {Linder}}\ and\ \bibinfo {author} {\bibfnamefont {J.~W.~A.}\ \bibnamefont
  {Robinson}},\ }\href {\doibase 10.1038/nphys3242} {\bibfield  {journal}
  {\bibinfo  {journal} {Nat. Phys.}\ }\textbf {\bibinfo {volume} {11}},\
  \bibinfo {pages} {307} (\bibinfo {year} {2015})}\BibitemShut {NoStop}%
\bibitem [{\citenamefont {Josephson}(1962)}]{1962JOS}%
  \BibitemOpen
  \bibfield  {author} {\bibinfo {author} {\bibfnamefont {B.}~\bibnamefont
  {Josephson}},\ }\href {\doibase 10.1016/0031-9163(62)91369-0} {\bibfield
  {journal} {\bibinfo  {journal} {Phys. Lett}\ }\textbf {\bibinfo {volume}
  {1}},\ \bibinfo {pages} {251} (\bibinfo {year} {1962})}\BibitemShut {NoStop}%
\bibitem [{\citenamefont {Bergen}\ \emph {et~al.}(2019)\citenamefont {Bergen},
  \citenamefont {Andersen}, \citenamefont {Bauer}, \citenamefont {Boy},
  \citenamefont {ter Brake}, \citenamefont {Brutsaert}, \citenamefont
  {B{\"{u}}hrer}, \citenamefont {Dhall{\'{e}}}, \citenamefont {Hansen},
  \citenamefont {ten Kate}, \citenamefont {Kellers}, \citenamefont {Krause},
  \citenamefont {Krooshoop}, \citenamefont {Kruse}, \citenamefont {Kylling},
  \citenamefont {Pilas}, \citenamefont {P{\"{u}}tz}, \citenamefont {Rebsdorf},
  \citenamefont {Reckhard}, \citenamefont {Seitz}, \citenamefont {Springer},
  \citenamefont {Song}, \citenamefont {Tzabar}, \citenamefont {Wessel},
  \citenamefont {Wiezoreck}, \citenamefont {Winkler},\ and\ \citenamefont
  {Yagotyntsev}}]{2019BER}%
  \BibitemOpen
  \bibfield  {author} {\bibinfo {author} {\bibfnamefont {A.}~\bibnamefont
  {Bergen}}, \bibinfo {author} {\bibfnamefont {R.}~\bibnamefont {Andersen}},
  \bibinfo {author} {\bibfnamefont {M.}~\bibnamefont {Bauer}}, \bibinfo
  {author} {\bibfnamefont {H.}~\bibnamefont {Boy}}, \bibinfo {author}
  {\bibfnamefont {M.}~\bibnamefont {ter Brake}}, \bibinfo {author}
  {\bibfnamefont {P.}~\bibnamefont {Brutsaert}}, \bibinfo {author}
  {\bibfnamefont {C.}~\bibnamefont {B{\"{u}}hrer}}, \bibinfo {author}
  {\bibfnamefont {M.}~\bibnamefont {Dhall{\'{e}}}}, \bibinfo {author}
  {\bibfnamefont {J.}~\bibnamefont {Hansen}}, \bibinfo {author} {\bibfnamefont
  {H.}~\bibnamefont {ten Kate}}, \bibinfo {author} {\bibfnamefont
  {J.}~\bibnamefont {Kellers}}, \bibinfo {author} {\bibfnamefont
  {J.}~\bibnamefont {Krause}}, \bibinfo {author} {\bibfnamefont
  {E.}~\bibnamefont {Krooshoop}}, \bibinfo {author} {\bibfnamefont
  {C.}~\bibnamefont {Kruse}}, \bibinfo {author} {\bibfnamefont
  {H.}~\bibnamefont {Kylling}}, \bibinfo {author} {\bibfnamefont
  {M.}~\bibnamefont {Pilas}}, \bibinfo {author} {\bibfnamefont
  {H.}~\bibnamefont {P{\"{u}}tz}}, \bibinfo {author} {\bibfnamefont
  {A.}~\bibnamefont {Rebsdorf}}, \bibinfo {author} {\bibfnamefont
  {M.}~\bibnamefont {Reckhard}}, \bibinfo {author} {\bibfnamefont
  {E.}~\bibnamefont {Seitz}}, \bibinfo {author} {\bibfnamefont
  {H.}~\bibnamefont {Springer}}, \bibinfo {author} {\bibfnamefont
  {X.}~\bibnamefont {Song}}, \bibinfo {author} {\bibfnamefont {N.}~\bibnamefont
  {Tzabar}}, \bibinfo {author} {\bibfnamefont {S.}~\bibnamefont {Wessel}},
  \bibinfo {author} {\bibfnamefont {J.}~\bibnamefont {Wiezoreck}}, \bibinfo
  {author} {\bibfnamefont {T.}~\bibnamefont {Winkler}}, \ and\ \bibinfo
  {author} {\bibfnamefont {K.}~\bibnamefont {Yagotyntsev}},\ }\href {\doibase
  10.1088/1361-6668/ab48d6} {\bibfield  {journal} {\bibinfo  {journal}
  {Supercond. Sci. Technol.}\ }\textbf {\bibinfo {volume} {32}},\ \bibinfo
  {pages} {125006} (\bibinfo {year} {2019})}\BibitemShut {NoStop}%
\bibitem [{\citenamefont {Islam}\ \emph {et~al.}(2014)\citenamefont {Islam},
  \citenamefont {Guo},\ and\ \citenamefont {Zhu}}]{2014ISL}%
  \BibitemOpen
  \bibfield  {author} {\bibinfo {author} {\bibfnamefont {M.~R.}\ \bibnamefont
  {Islam}}, \bibinfo {author} {\bibfnamefont {Y.}~\bibnamefont {Guo}}, \ and\
  \bibinfo {author} {\bibfnamefont {J.}~\bibnamefont {Zhu}},\ }\href {\doibase
  10.1016/j.rser.2014.01.085} {\bibfield  {journal} {\bibinfo  {journal}
  {Renew. Sust. Energ. Rev.}\ }\textbf {\bibinfo {volume} {33}},\ \bibinfo
  {pages} {161} (\bibinfo {year} {2014})}\BibitemShut {NoStop}%
\bibitem [{\citenamefont {Thomas}\ \emph {et~al.}(2016)\citenamefont {Thomas},
  \citenamefont {Marian}, \citenamefont {Chervyakov}, \citenamefont
  {St{\"{u}}ckrad}, \citenamefont {Salmieri},\ and\ \citenamefont
  {Rubbia}}]{2016THO}%
  \BibitemOpen
  \bibfield  {author} {\bibinfo {author} {\bibfnamefont {H.}~\bibnamefont
  {Thomas}}, \bibinfo {author} {\bibfnamefont {A.}~\bibnamefont {Marian}},
  \bibinfo {author} {\bibfnamefont {A.}~\bibnamefont {Chervyakov}}, \bibinfo
  {author} {\bibfnamefont {S.}~\bibnamefont {St{\"{u}}ckrad}}, \bibinfo
  {author} {\bibfnamefont {D.}~\bibnamefont {Salmieri}}, \ and\ \bibinfo
  {author} {\bibfnamefont {C.}~\bibnamefont {Rubbia}},\ }\href {\doibase
  10.1016/j.rser.2015.10.041} {\bibfield  {journal} {\bibinfo  {journal}
  {Renew. Sust. Energ. Rev.}\ }\textbf {\bibinfo {volume} {55}},\ \bibinfo
  {pages} {59} (\bibinfo {year} {2016})}\BibitemShut {NoStop}%
\bibitem [{\citenamefont {Tinkham}(1988)}]{1988TIN}%
  \BibitemOpen
  \bibfield  {author} {\bibinfo {author} {\bibfnamefont {M.}~\bibnamefont
  {Tinkham}},\ }\href {\doibase 10.1103/physrevlett.61.1658} {\bibfield
  {journal} {\bibinfo  {journal} {Phys. Rev. Lett.}\ }\textbf {\bibinfo
  {volume} {61}},\ \bibinfo {pages} {1658} (\bibinfo {year}
  {1988})}\BibitemShut {NoStop}%
\bibitem [{\citenamefont {Kalsi}(2011)}]{2011KAL}%
  \BibitemOpen
  \bibfield  {author} {\bibinfo {author} {\bibfnamefont {S.~S.}\ \bibnamefont
  {Kalsi}},\ }\href@noop {} {\emph {\bibinfo {title} {Applications of high
  temperature superconductors to electric power equipment}}}\ (\bibinfo
  {publisher} {John Wiley \& Sons},\ \bibinfo {year} {2011})\BibitemShut
  {NoStop}%
\bibitem [{\citenamefont {Flores-Livas}\ \emph {et~al.}(2020)\citenamefont
  {Flores-Livas}, \citenamefont {Boeri}, \citenamefont {Sanna}, \citenamefont
  {Profeta}, \citenamefont {Arita},\ and\ \citenamefont {Eremets}}]{2020FLO}%
  \BibitemOpen
  \bibfield  {author} {\bibinfo {author} {\bibfnamefont {J.~A.}\ \bibnamefont
  {Flores-Livas}}, \bibinfo {author} {\bibfnamefont {L.}~\bibnamefont {Boeri}},
  \bibinfo {author} {\bibfnamefont {A.}~\bibnamefont {Sanna}}, \bibinfo
  {author} {\bibfnamefont {G.}~\bibnamefont {Profeta}}, \bibinfo {author}
  {\bibfnamefont {R.}~\bibnamefont {Arita}}, \ and\ \bibinfo {author}
  {\bibfnamefont {M.}~\bibnamefont {Eremets}},\ }\href {\doibase
  10.1016/j.physrep.2020.02.003} {\bibfield  {journal} {\bibinfo  {journal}
  {Phys. Rep.}\ }\textbf {\bibinfo {volume} {856}},\ \bibinfo {pages} {1}
  (\bibinfo {year} {2020})}\BibitemShut {NoStop}%
\bibitem [{\citenamefont {Bardeen}(1955)}]{1955BAR}%
  \BibitemOpen
  \bibfield  {author} {\bibinfo {author} {\bibfnamefont {J.}~\bibnamefont
  {Bardeen}},\ }\href {\doibase 10.1103/physrev.97.1724} {\bibfield  {journal}
  {\bibinfo  {journal} {Phys. Rev.}\ }\textbf {\bibinfo {volume} {97}},\
  \bibinfo {pages} {1724} (\bibinfo {year} {1955})}\BibitemShut {NoStop}%
\bibitem [{\citenamefont {Cooper}(1956)}]{1956COO}%
  \BibitemOpen
  \bibfield  {author} {\bibinfo {author} {\bibfnamefont {L.~N.}\ \bibnamefont
  {Cooper}},\ }\href {\doibase 10.1103/physrev.104.1189} {\bibfield  {journal}
  {\bibinfo  {journal} {Phys. Rev.}\ }\textbf {\bibinfo {volume} {104}},\
  \bibinfo {pages} {1189} (\bibinfo {year} {1956})}\BibitemShut {NoStop}%
\bibitem [{\citenamefont {Bardeen}\ \emph
  {et~al.}(1957{\natexlab{a}})\citenamefont {Bardeen}, \citenamefont {Cooper},\
  and\ \citenamefont {Schrieffer}}]{1957BARa}%
  \BibitemOpen
  \bibfield  {author} {\bibinfo {author} {\bibfnamefont {J.}~\bibnamefont
  {Bardeen}}, \bibinfo {author} {\bibfnamefont {L.~N.}\ \bibnamefont {Cooper}},
  \ and\ \bibinfo {author} {\bibfnamefont {J.~R.}\ \bibnamefont {Schrieffer}},\
  }\href {\doibase 10.1103/physrev.106.162} {\bibfield  {journal} {\bibinfo
  {journal} {Phys. Rev.}\ }\textbf {\bibinfo {volume} {106}},\ \bibinfo {pages}
  {162} (\bibinfo {year} {1957}{\natexlab{a}})}\BibitemShut {NoStop}%
\bibitem [{\citenamefont {Bardeen}\ \emph
  {et~al.}(1957{\natexlab{b}})\citenamefont {Bardeen}, \citenamefont {Cooper},\
  and\ \citenamefont {Schrieffer}}]{1957BARb}%
  \BibitemOpen
  \bibfield  {author} {\bibinfo {author} {\bibfnamefont {J.}~\bibnamefont
  {Bardeen}}, \bibinfo {author} {\bibfnamefont {L.~N.}\ \bibnamefont {Cooper}},
  \ and\ \bibinfo {author} {\bibfnamefont {J.~R.}\ \bibnamefont {Schrieffer}},\
  }\href {\doibase 10.1103/physrev.108.1175} {\bibfield  {journal} {\bibinfo
  {journal} {Phys. Rev.}\ }\textbf {\bibinfo {volume} {108}},\ \bibinfo {pages}
  {1175} (\bibinfo {year} {1957}{\natexlab{b}})}\BibitemShut {NoStop}%
\bibitem [{\citenamefont {Ashcroft}(1968)}]{1968ASH}%
  \BibitemOpen
  \bibfield  {author} {\bibinfo {author} {\bibfnamefont {N.~W.}\ \bibnamefont
  {Ashcroft}},\ }\href {\doibase 10.1103/physrevlett.21.1748} {\bibfield
  {journal} {\bibinfo  {journal} {Phys. Rev. Lett.}\ }\textbf {\bibinfo
  {volume} {21}},\ \bibinfo {pages} {1748} (\bibinfo {year}
  {1968})}\BibitemShut {NoStop}%
\bibitem [{\citenamefont {Wigner}\ and\ \citenamefont
  {Huntington}(1935)}]{1935WIG}%
  \BibitemOpen
  \bibfield  {author} {\bibinfo {author} {\bibfnamefont {E.}~\bibnamefont
  {Wigner}}\ and\ \bibinfo {author} {\bibfnamefont {H.~B.}\ \bibnamefont
  {Huntington}},\ }\href {\doibase 10.1063/1.1749590} {\bibfield  {journal}
  {\bibinfo  {journal} {J. Chem. Phys.}\ }\textbf {\bibinfo {volume} {3}},\
  \bibinfo {pages} {764} (\bibinfo {year} {1935})}\BibitemShut {NoStop}%
\bibitem [{\citenamefont {Loubeyre}\ \emph {et~al.}(1996)\citenamefont
  {Loubeyre}, \citenamefont {LeToullec}, \citenamefont {Hausermann},
  \citenamefont {Hanfland}, \citenamefont {Hemley}, \citenamefont {Mao},\ and\
  \citenamefont {Finger}}]{1996LOU}%
  \BibitemOpen
  \bibfield  {author} {\bibinfo {author} {\bibfnamefont {P.}~\bibnamefont
  {Loubeyre}}, \bibinfo {author} {\bibfnamefont {R.}~\bibnamefont {LeToullec}},
  \bibinfo {author} {\bibfnamefont {D.}~\bibnamefont {Hausermann}}, \bibinfo
  {author} {\bibfnamefont {M.}~\bibnamefont {Hanfland}}, \bibinfo {author}
  {\bibfnamefont {R.~J.}\ \bibnamefont {Hemley}}, \bibinfo {author}
  {\bibfnamefont {H.~K.}\ \bibnamefont {Mao}}, \ and\ \bibinfo {author}
  {\bibfnamefont {L.~W.}\ \bibnamefont {Finger}},\ }\href {\doibase
  10.1038/383702a0} {\bibfield  {journal} {\bibinfo  {journal} {Nature}\
  }\textbf {\bibinfo {volume} {383}},\ \bibinfo {pages} {702} (\bibinfo {year}
  {1996})}\BibitemShut {NoStop}%
\bibitem [{\citenamefont {McMinis}\ \emph {et~al.}(2015)\citenamefont
  {McMinis}, \citenamefont {Clay}, \citenamefont {Lee},\ and\ \citenamefont
  {Morales}}]{2015MCM}%
  \BibitemOpen
  \bibfield  {author} {\bibinfo {author} {\bibfnamefont {J.}~\bibnamefont
  {McMinis}}, \bibinfo {author} {\bibfnamefont {R.~C.}\ \bibnamefont {Clay}},
  \bibinfo {author} {\bibfnamefont {D.}~\bibnamefont {Lee}}, \ and\ \bibinfo
  {author} {\bibfnamefont {M.~A.}\ \bibnamefont {Morales}},\ }\href {\doibase
  10.1103/physrevlett.114.105305} {\bibfield  {journal} {\bibinfo  {journal}
  {Phys. Rev. Lett.}\ }\textbf {\bibinfo {volume} {114}},\ \bibinfo {pages}
  {105305} (\bibinfo {year} {2015})}\BibitemShut {NoStop}%
\bibitem [{\citenamefont {Eremets}\ and\ \citenamefont
  {Troyan}(2011)}]{2011ERE}%
  \BibitemOpen
  \bibfield  {author} {\bibinfo {author} {\bibfnamefont {M.~I.}\ \bibnamefont
  {Eremets}}\ and\ \bibinfo {author} {\bibfnamefont {I.~A.}\ \bibnamefont
  {Troyan}},\ }\href {\doibase 10.1038/nmat3175} {\bibfield  {journal}
  {\bibinfo  {journal} {Nat. Mater.}\ }\textbf {\bibinfo {volume} {10}},\
  \bibinfo {pages} {927} (\bibinfo {year} {2011})}\BibitemShut {NoStop}%
\bibitem [{\citenamefont {Dalladay-Simpson}\ \emph {et~al.}(2016)\citenamefont
  {Dalladay-Simpson}, \citenamefont {Howie},\ and\ \citenamefont
  {Gregoryanz}}]{2016DAL}%
  \BibitemOpen
  \bibfield  {author} {\bibinfo {author} {\bibfnamefont {P.}~\bibnamefont
  {Dalladay-Simpson}}, \bibinfo {author} {\bibfnamefont {R.~T.}\ \bibnamefont
  {Howie}}, \ and\ \bibinfo {author} {\bibfnamefont {E.}~\bibnamefont
  {Gregoryanz}},\ }\href {\doibase 10.1038/nature16164} {\bibfield  {journal}
  {\bibinfo  {journal} {Nature}\ }\textbf {\bibinfo {volume} {529}},\ \bibinfo
  {pages} {63} (\bibinfo {year} {2016})}\BibitemShut {NoStop}%
\bibitem [{\citenamefont {Dias}\ and\ \citenamefont {Silvera}(2017)}]{2017DIA}%
  \BibitemOpen
  \bibfield  {author} {\bibinfo {author} {\bibfnamefont {R.~P.}\ \bibnamefont
  {Dias}}\ and\ \bibinfo {author} {\bibfnamefont {I.~F.}\ \bibnamefont
  {Silvera}},\ }\href {\doibase 10.1126/science.aal1579} {\bibfield  {journal}
  {\bibinfo  {journal} {Science}\ }\textbf {\bibinfo {volume} {355}},\ \bibinfo
  {pages} {715} (\bibinfo {year} {2017})}\BibitemShut {NoStop}%
\bibitem [{\citenamefont {Castelvecchi}(2017)}]{2017CAS}%
  \BibitemOpen
  \bibfield  {author} {\bibinfo {author} {\bibfnamefont {D.}~\bibnamefont
  {Castelvecchi}},\ }\href {\doibase 10.1038/nature.2017.21379} {\bibfield
  {journal} {\bibinfo  {journal} {Nature}\ }\textbf {\bibinfo {volume} {542}},\
  \bibinfo {pages} {17} (\bibinfo {year} {2017})}\BibitemShut {NoStop}%
\bibitem [{\citenamefont {Celliers}\ \emph {et~al.}(2018)\citenamefont
  {Celliers}, \citenamefont {Millot}, \citenamefont {Brygoo}, \citenamefont
  {McWilliams}, \citenamefont {Fratanduono}, \citenamefont {Rygg},
  \citenamefont {Goncharov}, \citenamefont {Loubeyre}, \citenamefont {Eggert},
  \citenamefont {Peterson}, \citenamefont {Meezan}, \citenamefont {Pape},
  \citenamefont {Collins}, \citenamefont {Jeanloz},\ and\ \citenamefont
  {Hemley}}]{2018CEL}%
  \BibitemOpen
  \bibfield  {author} {\bibinfo {author} {\bibfnamefont {P.~M.}\ \bibnamefont
  {Celliers}}, \bibinfo {author} {\bibfnamefont {M.}~\bibnamefont {Millot}},
  \bibinfo {author} {\bibfnamefont {S.}~\bibnamefont {Brygoo}}, \bibinfo
  {author} {\bibfnamefont {R.~S.}\ \bibnamefont {McWilliams}}, \bibinfo
  {author} {\bibfnamefont {D.~E.}\ \bibnamefont {Fratanduono}}, \bibinfo
  {author} {\bibfnamefont {J.~R.}\ \bibnamefont {Rygg}}, \bibinfo {author}
  {\bibfnamefont {A.~F.}\ \bibnamefont {Goncharov}}, \bibinfo {author}
  {\bibfnamefont {P.}~\bibnamefont {Loubeyre}}, \bibinfo {author}
  {\bibfnamefont {J.~H.}\ \bibnamefont {Eggert}}, \bibinfo {author}
  {\bibfnamefont {J.~L.}\ \bibnamefont {Peterson}}, \bibinfo {author}
  {\bibfnamefont {N.~B.}\ \bibnamefont {Meezan}}, \bibinfo {author}
  {\bibfnamefont {S.~L.}\ \bibnamefont {Pape}}, \bibinfo {author}
  {\bibfnamefont {G.~W.}\ \bibnamefont {Collins}}, \bibinfo {author}
  {\bibfnamefont {R.}~\bibnamefont {Jeanloz}}, \ and\ \bibinfo {author}
  {\bibfnamefont {R.~J.}\ \bibnamefont {Hemley}},\ }\href {\doibase
  10.1126/science.aat0970} {\bibfield  {journal} {\bibinfo  {journal}
  {Science}\ }\textbf {\bibinfo {volume} {361}},\ \bibinfo {pages} {677}
  (\bibinfo {year} {2018})}\BibitemShut {NoStop}%
\bibitem [{\citenamefont {Ashcroft}(2004)}]{2004ASH}%
  \BibitemOpen
  \bibfield  {author} {\bibinfo {author} {\bibfnamefont {N.~W.}\ \bibnamefont
  {Ashcroft}},\ }\href {\doibase 10.1103/physrevlett.92.187002} {\bibfield
  {journal} {\bibinfo  {journal} {Phys. Rev. Lett.}\ }\textbf {\bibinfo
  {volume} {92}},\ \bibinfo {pages} {187002} (\bibinfo {year}
  {2004})}\BibitemShut {NoStop}%
\bibitem [{\citenamefont {Semenok}\ \emph {et~al.}(2020)\citenamefont
  {Semenok}, \citenamefont {Kruglov}, \citenamefont {Savkin}, \citenamefont
  {Kvashnin},\ and\ \citenamefont {Oganov}}]{2020SEM}%
  \BibitemOpen
  \bibfield  {author} {\bibinfo {author} {\bibfnamefont {D.~V.}\ \bibnamefont
  {Semenok}}, \bibinfo {author} {\bibfnamefont {I.~A.}\ \bibnamefont
  {Kruglov}}, \bibinfo {author} {\bibfnamefont {I.~A.}\ \bibnamefont {Savkin}},
  \bibinfo {author} {\bibfnamefont {A.~G.}\ \bibnamefont {Kvashnin}}, \ and\
  \bibinfo {author} {\bibfnamefont {A.~R.}\ \bibnamefont {Oganov}},\ }\href
  {\doibase 10.1016/j.cossms.2020.100808} {\bibfield  {journal} {\bibinfo
  {journal} {Curr. Opin. Solid State Mater. Sci.}\ }\textbf {\bibinfo {volume}
  {24}},\ \bibinfo {pages} {100808} (\bibinfo {year} {2020})}\BibitemShut
  {NoStop}%
\bibitem [{\citenamefont {Peng}\ \emph {et~al.}(2017)\citenamefont {Peng},
  \citenamefont {Sun}, \citenamefont {Pickard}, \citenamefont {Needs},
  \citenamefont {Wu},\ and\ \citenamefont {Ma}}]{2017PEN}%
  \BibitemOpen
  \bibfield  {author} {\bibinfo {author} {\bibfnamefont {F.}~\bibnamefont
  {Peng}}, \bibinfo {author} {\bibfnamefont {Y.}~\bibnamefont {Sun}}, \bibinfo
  {author} {\bibfnamefont {C.~J.}\ \bibnamefont {Pickard}}, \bibinfo {author}
  {\bibfnamefont {R.~J.}\ \bibnamefont {Needs}}, \bibinfo {author}
  {\bibfnamefont {Q.}~\bibnamefont {Wu}}, \ and\ \bibinfo {author}
  {\bibfnamefont {Y.}~\bibnamefont {Ma}},\ }\href {\doibase
  10.1103/physrevlett.119.107001} {\bibfield  {journal} {\bibinfo  {journal}
  {Phys. Rev. Lett.}\ }\textbf {\bibinfo {volume} {119}},\ \bibinfo {pages}
  {107001} (\bibinfo {year} {2017})}\BibitemShut {NoStop}%
\bibitem [{\citenamefont {Zhang}\ \emph {et~al.}(2020)\citenamefont {Zhang},
  \citenamefont {McMahon}, \citenamefont {Oganov}, \citenamefont {Li},
  \citenamefont {Dong}, \citenamefont {Dong},\ and\ \citenamefont
  {Wang}}]{2020ZHA}%
  \BibitemOpen
  \bibfield  {author} {\bibinfo {author} {\bibfnamefont {J.}~\bibnamefont
  {Zhang}}, \bibinfo {author} {\bibfnamefont {J.~M.}\ \bibnamefont {McMahon}},
  \bibinfo {author} {\bibfnamefont {A.~R.}\ \bibnamefont {Oganov}}, \bibinfo
  {author} {\bibfnamefont {X.}~\bibnamefont {Li}}, \bibinfo {author}
  {\bibfnamefont {X.}~\bibnamefont {Dong}}, \bibinfo {author} {\bibfnamefont
  {H.}~\bibnamefont {Dong}}, \ and\ \bibinfo {author} {\bibfnamefont
  {S.}~\bibnamefont {Wang}},\ }\href {\doibase 10.1103/physrevb.101.134108}
  {\bibfield  {journal} {\bibinfo  {journal} {Phys. Rev. B}\ }\textbf {\bibinfo
  {volume} {101}},\ \bibinfo {pages} {134108} (\bibinfo {year}
  {2020})}\BibitemShut {NoStop}%
\bibitem [{\citenamefont {Li}\ and\ \citenamefont {Peng}(2017)}]{2017LI}%
  \BibitemOpen
  \bibfield  {author} {\bibinfo {author} {\bibfnamefont {X.}~\bibnamefont
  {Li}}\ and\ \bibinfo {author} {\bibfnamefont {F.}~\bibnamefont {Peng}},\
  }\href {\doibase 10.1021/acs.inorgchem.7b01686} {\bibfield  {journal}
  {\bibinfo  {journal} {Inorg. Chem.}\ }\textbf {\bibinfo {volume} {56}},\
  \bibinfo {pages} {13759} (\bibinfo {year} {2017})}\BibitemShut {NoStop}%
\bibitem [{\citenamefont {Yu}\ \emph {et~al.}(2015)\citenamefont {Yu},
  \citenamefont {Jia}, \citenamefont {Frapper}, \citenamefont {Li},
  \citenamefont {Oganov}, \citenamefont {Zeng},\ and\ \citenamefont
  {Zhang}}]{2015YU}%
  \BibitemOpen
  \bibfield  {author} {\bibinfo {author} {\bibfnamefont {S.}~\bibnamefont
  {Yu}}, \bibinfo {author} {\bibfnamefont {X.}~\bibnamefont {Jia}}, \bibinfo
  {author} {\bibfnamefont {G.}~\bibnamefont {Frapper}}, \bibinfo {author}
  {\bibfnamefont {D.}~\bibnamefont {Li}}, \bibinfo {author} {\bibfnamefont
  {A.~R.}\ \bibnamefont {Oganov}}, \bibinfo {author} {\bibfnamefont
  {Q.}~\bibnamefont {Zeng}}, \ and\ \bibinfo {author} {\bibfnamefont
  {L.}~\bibnamefont {Zhang}},\ }\href {\doibase 10.1038/srep17764} {\bibfield
  {journal} {\bibinfo  {journal} {Sci. Rep.}\ }\textbf {\bibinfo {volume}
  {5}},\ \bibinfo {pages} {1} (\bibinfo {year} {2015})}\BibitemShut {NoStop}%
\bibitem [{\citenamefont {Einaga}\ \emph {et~al.}(2016)\citenamefont {Einaga},
  \citenamefont {Sakata}, \citenamefont {Ishikawa}, \citenamefont {Shimizu},
  \citenamefont {Eremets}, \citenamefont {Drozdov}, \citenamefont {Troyan},
  \citenamefont {Hirao},\ and\ \citenamefont {Ohishi}}]{2016EIN}%
  \BibitemOpen
  \bibfield  {author} {\bibinfo {author} {\bibfnamefont {M.}~\bibnamefont
  {Einaga}}, \bibinfo {author} {\bibfnamefont {M.}~\bibnamefont {Sakata}},
  \bibinfo {author} {\bibfnamefont {T.}~\bibnamefont {Ishikawa}}, \bibinfo
  {author} {\bibfnamefont {K.}~\bibnamefont {Shimizu}}, \bibinfo {author}
  {\bibfnamefont {M.~I.}\ \bibnamefont {Eremets}}, \bibinfo {author}
  {\bibfnamefont {A.~P.}\ \bibnamefont {Drozdov}}, \bibinfo {author}
  {\bibfnamefont {I.~A.}\ \bibnamefont {Troyan}}, \bibinfo {author}
  {\bibfnamefont {N.}~\bibnamefont {Hirao}}, \ and\ \bibinfo {author}
  {\bibfnamefont {Y.}~\bibnamefont {Ohishi}},\ }\href {\doibase
  10.1038/nphys3760} {\bibfield  {journal} {\bibinfo  {journal} {Nat. Phys.}\
  }\textbf {\bibinfo {volume} {12}},\ \bibinfo {pages} {835} (\bibinfo {year}
  {2016})}\BibitemShut {NoStop}%
\bibitem [{\citenamefont {Drozdov}\ \emph {et~al.}(2015)\citenamefont
  {Drozdov}, \citenamefont {Eremets}, \citenamefont {Troyan}, \citenamefont
  {Ksenofontov},\ and\ \citenamefont {Shylin}}]{2015DRO}%
  \BibitemOpen
  \bibfield  {author} {\bibinfo {author} {\bibfnamefont {A.~P.}\ \bibnamefont
  {Drozdov}}, \bibinfo {author} {\bibfnamefont {M.~I.}\ \bibnamefont
  {Eremets}}, \bibinfo {author} {\bibfnamefont {I.~A.}\ \bibnamefont {Troyan}},
  \bibinfo {author} {\bibfnamefont {V.}~\bibnamefont {Ksenofontov}}, \ and\
  \bibinfo {author} {\bibfnamefont {S.~I.}\ \bibnamefont {Shylin}},\ }\href
  {\doibase 10.1038/nature14964} {\bibfield  {journal} {\bibinfo  {journal}
  {Nature}\ }\textbf {\bibinfo {volume} {525}},\ \bibinfo {pages} {73}
  (\bibinfo {year} {2015})}\BibitemShut {NoStop}%
\bibitem [{\citenamefont {Drozdov}\ \emph {et~al.}(2019)\citenamefont
  {Drozdov}, \citenamefont {Kong}, \citenamefont {Minkov}, \citenamefont
  {Besedin}, \citenamefont {Kuzovnikov}, \citenamefont {Mozaffari},
  \citenamefont {Balicas}, \citenamefont {Balakirev}, \citenamefont {Graf},
  \citenamefont {Prakapenka}, \citenamefont {Greenberg}, \citenamefont
  {Knyazev}, \citenamefont {Tkacz},\ and\ \citenamefont {Eremets}}]{2019DRO}%
  \BibitemOpen
  \bibfield  {author} {\bibinfo {author} {\bibfnamefont {A.~P.}\ \bibnamefont
  {Drozdov}}, \bibinfo {author} {\bibfnamefont {P.~P.}\ \bibnamefont {Kong}},
  \bibinfo {author} {\bibfnamefont {V.~S.}\ \bibnamefont {Minkov}}, \bibinfo
  {author} {\bibfnamefont {S.~P.}\ \bibnamefont {Besedin}}, \bibinfo {author}
  {\bibfnamefont {M.~A.}\ \bibnamefont {Kuzovnikov}}, \bibinfo {author}
  {\bibfnamefont {S.}~\bibnamefont {Mozaffari}}, \bibinfo {author}
  {\bibfnamefont {L.}~\bibnamefont {Balicas}}, \bibinfo {author} {\bibfnamefont
  {F.~F.}\ \bibnamefont {Balakirev}}, \bibinfo {author} {\bibfnamefont {D.~E.}\
  \bibnamefont {Graf}}, \bibinfo {author} {\bibfnamefont {V.~B.}\ \bibnamefont
  {Prakapenka}}, \bibinfo {author} {\bibfnamefont {E.}~\bibnamefont
  {Greenberg}}, \bibinfo {author} {\bibfnamefont {D.~A.}\ \bibnamefont
  {Knyazev}}, \bibinfo {author} {\bibfnamefont {M.}~\bibnamefont {Tkacz}}, \
  and\ \bibinfo {author} {\bibfnamefont {M.~I.}\ \bibnamefont {Eremets}},\
  }\href {\doibase 10.1038/s41586-019-1201-8} {\bibfield  {journal} {\bibinfo
  {journal} {Nature}\ }\textbf {\bibinfo {volume} {569}},\ \bibinfo {pages}
  {528} (\bibinfo {year} {2019})}\BibitemShut {NoStop}%
\bibitem [{\citenamefont {Salke}\ \emph {et~al.}(2019)\citenamefont {Salke},
  \citenamefont {Esfahani}, \citenamefont {Zhang}, \citenamefont {Kruglov},
  \citenamefont {Zhou}, \citenamefont {Wang}, \citenamefont {Greenberg},
  \citenamefont {Prakapenka}, \citenamefont {Liu}, \citenamefont {Oganov},\
  and\ \citenamefont {Lin}}]{2019SAL}%
  \BibitemOpen
  \bibfield  {author} {\bibinfo {author} {\bibfnamefont {N.~P.}\ \bibnamefont
  {Salke}}, \bibinfo {author} {\bibfnamefont {M.~M.~D.}\ \bibnamefont
  {Esfahani}}, \bibinfo {author} {\bibfnamefont {Y.}~\bibnamefont {Zhang}},
  \bibinfo {author} {\bibfnamefont {I.~A.}\ \bibnamefont {Kruglov}}, \bibinfo
  {author} {\bibfnamefont {J.}~\bibnamefont {Zhou}}, \bibinfo {author}
  {\bibfnamefont {Y.}~\bibnamefont {Wang}}, \bibinfo {author} {\bibfnamefont
  {E.}~\bibnamefont {Greenberg}}, \bibinfo {author} {\bibfnamefont {V.~B.}\
  \bibnamefont {Prakapenka}}, \bibinfo {author} {\bibfnamefont
  {J.}~\bibnamefont {Liu}}, \bibinfo {author} {\bibfnamefont {A.~R.}\
  \bibnamefont {Oganov}}, \ and\ \bibinfo {author} {\bibfnamefont {J.-F.}\
  \bibnamefont {Lin}},\ }\href {\doibase 10.1038/s41467-019-12326-y} {\bibfield
   {journal} {\bibinfo  {journal} {Nat. Commun.}\ }\textbf {\bibinfo {volume}
  {10}},\ \bibinfo {pages} {1} (\bibinfo {year} {2019})}\BibitemShut {NoStop}%
\bibitem [{\citenamefont {Chen}\ \emph
  {et~al.}(2021{\natexlab{a}})\citenamefont {Chen}, \citenamefont {Semenok},
  \citenamefont {Kvashnin}, \citenamefont {Huang}, \citenamefont {Kruglov},
  \citenamefont {Galasso}, \citenamefont {Song}, \citenamefont {Duan},
  \citenamefont {Goncharov}, \citenamefont {Prakapenka}, \citenamefont
  {Oganov},\ and\ \citenamefont {Cui}}]{2021CHEa}%
  \BibitemOpen
  \bibfield  {author} {\bibinfo {author} {\bibfnamefont {W.}~\bibnamefont
  {Chen}}, \bibinfo {author} {\bibfnamefont {D.~V.}\ \bibnamefont {Semenok}},
  \bibinfo {author} {\bibfnamefont {A.~G.}\ \bibnamefont {Kvashnin}}, \bibinfo
  {author} {\bibfnamefont {X.}~\bibnamefont {Huang}}, \bibinfo {author}
  {\bibfnamefont {I.~A.}\ \bibnamefont {Kruglov}}, \bibinfo {author}
  {\bibfnamefont {M.}~\bibnamefont {Galasso}}, \bibinfo {author} {\bibfnamefont
  {H.}~\bibnamefont {Song}}, \bibinfo {author} {\bibfnamefont {D.}~\bibnamefont
  {Duan}}, \bibinfo {author} {\bibfnamefont {A.~F.}\ \bibnamefont {Goncharov}},
  \bibinfo {author} {\bibfnamefont {V.~B.}\ \bibnamefont {Prakapenka}},
  \bibinfo {author} {\bibfnamefont {A.~R.}\ \bibnamefont {Oganov}}, \ and\
  \bibinfo {author} {\bibfnamefont {T.}~\bibnamefont {Cui}},\ }\href {\doibase
  10.1038/s41467-020-20103-5} {\bibfield  {journal} {\bibinfo  {journal} {Nat.
  Commun.}\ }\textbf {\bibinfo {volume} {12}},\ \bibinfo {pages} {1} (\bibinfo
  {year} {2021}{\natexlab{a}})}\BibitemShut {NoStop}%
\bibitem [{\citenamefont {Shao}\ \emph
  {et~al.}(2021{\natexlab{a}})\citenamefont {Shao}, \citenamefont {Chen},
  \citenamefont {Zhang}, \citenamefont {Huang},\ and\ \citenamefont
  {Cui}}]{2021SHAa}%
  \BibitemOpen
  \bibfield  {author} {\bibinfo {author} {\bibfnamefont {M.}~\bibnamefont
  {Shao}}, \bibinfo {author} {\bibfnamefont {W.}~\bibnamefont {Chen}}, \bibinfo
  {author} {\bibfnamefont {K.}~\bibnamefont {Zhang}}, \bibinfo {author}
  {\bibfnamefont {X.}~\bibnamefont {Huang}}, \ and\ \bibinfo {author}
  {\bibfnamefont {T.}~\bibnamefont {Cui}},\ }\href {\doibase
  10.1103/physrevb.104.174509} {\bibfield  {journal} {\bibinfo  {journal}
  {Phys. Rev. B}\ }\textbf {\bibinfo {volume} {104}},\ \bibinfo {pages}
  {174509} (\bibinfo {year} {2021}{\natexlab{a}})}\BibitemShut {NoStop}%
\bibitem [{\citenamefont {Shao}\ \emph
  {et~al.}(2021{\natexlab{b}})\citenamefont {Shao}, \citenamefont {Chen},
  \citenamefont {Chen}, \citenamefont {Zhang}, \citenamefont {Huang},\ and\
  \citenamefont {Cui}}]{2021SHAb}%
  \BibitemOpen
  \bibfield  {author} {\bibinfo {author} {\bibfnamefont {M.}~\bibnamefont
  {Shao}}, \bibinfo {author} {\bibfnamefont {S.}~\bibnamefont {Chen}}, \bibinfo
  {author} {\bibfnamefont {W.}~\bibnamefont {Chen}}, \bibinfo {author}
  {\bibfnamefont {K.}~\bibnamefont {Zhang}}, \bibinfo {author} {\bibfnamefont
  {X.}~\bibnamefont {Huang}}, \ and\ \bibinfo {author} {\bibfnamefont
  {T.}~\bibnamefont {Cui}},\ }\href {\doibase 10.1021/acs.inorgchem.1c01960}
  {\bibfield  {journal} {\bibinfo  {journal} {Inorg. Chem.}\ }\textbf {\bibinfo
  {volume} {60}},\ \bibinfo {pages} {15330} (\bibinfo {year}
  {2021}{\natexlab{b}})}\BibitemShut {NoStop}%
\bibitem [{\citenamefont {Troyan}\ \emph {et~al.}(2021)\citenamefont {Troyan},
  \citenamefont {Semenok}, \citenamefont {Kvashnin}, \citenamefont {Sadakov},
  \citenamefont {Sobolevskiy}, \citenamefont {Pudalov}, \citenamefont
  {Ivanova}, \citenamefont {Prakapenka}, \citenamefont {Greenberg},
  \citenamefont {Gavriliuk}, \citenamefont {Lyubutin}, \citenamefont
  {Struzhkin}, \citenamefont {Bergara}, \citenamefont {Errea}, \citenamefont
  {Bianco}, \citenamefont {Calandra}, \citenamefont {Mauri}, \citenamefont
  {Monacelli}, \citenamefont {Akashi},\ and\ \citenamefont {Oganov}}]{2021TRO}%
  \BibitemOpen
  \bibfield  {author} {\bibinfo {author} {\bibfnamefont {I.~A.}\ \bibnamefont
  {Troyan}}, \bibinfo {author} {\bibfnamefont {D.~V.}\ \bibnamefont {Semenok}},
  \bibinfo {author} {\bibfnamefont {A.~G.}\ \bibnamefont {Kvashnin}}, \bibinfo
  {author} {\bibfnamefont {A.~V.}\ \bibnamefont {Sadakov}}, \bibinfo {author}
  {\bibfnamefont {O.~A.}\ \bibnamefont {Sobolevskiy}}, \bibinfo {author}
  {\bibfnamefont {V.~M.}\ \bibnamefont {Pudalov}}, \bibinfo {author}
  {\bibfnamefont {A.~G.}\ \bibnamefont {Ivanova}}, \bibinfo {author}
  {\bibfnamefont {V.~B.}\ \bibnamefont {Prakapenka}}, \bibinfo {author}
  {\bibfnamefont {E.}~\bibnamefont {Greenberg}}, \bibinfo {author}
  {\bibfnamefont {A.~G.}\ \bibnamefont {Gavriliuk}}, \bibinfo {author}
  {\bibfnamefont {I.~S.}\ \bibnamefont {Lyubutin}}, \bibinfo {author}
  {\bibfnamefont {V.~V.}\ \bibnamefont {Struzhkin}}, \bibinfo {author}
  {\bibfnamefont {A.}~\bibnamefont {Bergara}}, \bibinfo {author} {\bibfnamefont
  {I.}~\bibnamefont {Errea}}, \bibinfo {author} {\bibfnamefont
  {R.}~\bibnamefont {Bianco}}, \bibinfo {author} {\bibfnamefont
  {M.}~\bibnamefont {Calandra}}, \bibinfo {author} {\bibfnamefont
  {F.}~\bibnamefont {Mauri}}, \bibinfo {author} {\bibfnamefont
  {L.}~\bibnamefont {Monacelli}}, \bibinfo {author} {\bibfnamefont
  {R.}~\bibnamefont {Akashi}}, \ and\ \bibinfo {author} {\bibfnamefont {A.~R.}\
  \bibnamefont {Oganov}},\ }\href {\doibase 10.1002/adma.202006832} {\bibfield
  {journal} {\bibinfo  {journal} {Adv. Mater.}\ }\textbf {\bibinfo {volume}
  {33}},\ \bibinfo {pages} {2006832} (\bibinfo {year} {2021})}\BibitemShut
  {NoStop}%
\bibitem [{\citenamefont {Chen}\ \emph
  {et~al.}(2021{\natexlab{b}})\citenamefont {Chen}, \citenamefont {Semenok},
  \citenamefont {Huang}, \citenamefont {Shu}, \citenamefont {Li}, \citenamefont
  {Duan}, \citenamefont {Cui},\ and\ \citenamefont {Oganov}}]{2021CHEb}%
  \BibitemOpen
  \bibfield  {author} {\bibinfo {author} {\bibfnamefont {W.}~\bibnamefont
  {Chen}}, \bibinfo {author} {\bibfnamefont {D.}~\bibnamefont {Semenok}},
  \bibinfo {author} {\bibfnamefont {X.}~\bibnamefont {Huang}}, \bibinfo
  {author} {\bibfnamefont {H.}~\bibnamefont {Shu}}, \bibinfo {author}
  {\bibfnamefont {X.}~\bibnamefont {Li}}, \bibinfo {author} {\bibfnamefont
  {D.}~\bibnamefont {Duan}}, \bibinfo {author} {\bibfnamefont {T.}~\bibnamefont
  {Cui}}, \ and\ \bibinfo {author} {\bibfnamefont {A.}~\bibnamefont {Oganov}},\
  }\href {\doibase 10.1103/physrevlett.127.117001} {\bibfield  {journal}
  {\bibinfo  {journal} {Phys. Rev. Lett.}\ }\textbf {\bibinfo {volume} {127}},\
  \bibinfo {pages} {117001} (\bibinfo {year} {2021}{\natexlab{b}})}\BibitemShut
  {NoStop}%
\bibitem [{\citenamefont {Kong}\ \emph {et~al.}(2021)\citenamefont {Kong},
  \citenamefont {Minkov}, \citenamefont {Kuzovnikov}, \citenamefont {Drozdov},
  \citenamefont {Besedin}, \citenamefont {Mozaffari}, \citenamefont {Balicas},
  \citenamefont {Balakirev}, \citenamefont {Prakapenka}, \citenamefont
  {Chariton}, \citenamefont {Knyazev}, \citenamefont {Greenberg},\ and\
  \citenamefont {Eremets}}]{2021KON}%
  \BibitemOpen
  \bibfield  {author} {\bibinfo {author} {\bibfnamefont {P.}~\bibnamefont
  {Kong}}, \bibinfo {author} {\bibfnamefont {V.~S.}\ \bibnamefont {Minkov}},
  \bibinfo {author} {\bibfnamefont {M.~A.}\ \bibnamefont {Kuzovnikov}},
  \bibinfo {author} {\bibfnamefont {A.~P.}\ \bibnamefont {Drozdov}}, \bibinfo
  {author} {\bibfnamefont {S.~P.}\ \bibnamefont {Besedin}}, \bibinfo {author}
  {\bibfnamefont {S.}~\bibnamefont {Mozaffari}}, \bibinfo {author}
  {\bibfnamefont {L.}~\bibnamefont {Balicas}}, \bibinfo {author} {\bibfnamefont
  {F.~F.}\ \bibnamefont {Balakirev}}, \bibinfo {author} {\bibfnamefont {V.~B.}\
  \bibnamefont {Prakapenka}}, \bibinfo {author} {\bibfnamefont
  {S.}~\bibnamefont {Chariton}}, \bibinfo {author} {\bibfnamefont {D.~A.}\
  \bibnamefont {Knyazev}}, \bibinfo {author} {\bibfnamefont {E.}~\bibnamefont
  {Greenberg}}, \ and\ \bibinfo {author} {\bibfnamefont {M.~I.}\ \bibnamefont
  {Eremets}},\ }\href {\doibase 10.1038/s41467-021-25372-2} {\bibfield
  {journal} {\bibinfo  {journal} {Nat. Commun.}\ }\textbf {\bibinfo {volume}
  {12}},\ \bibinfo {pages} {1} (\bibinfo {year} {2021})}\BibitemShut {NoStop}%
\bibitem [{\citenamefont {Li}\ \emph {et~al.}(2014)\citenamefont {Li},
  \citenamefont {Hao}, \citenamefont {Liu}, \citenamefont {Li},\ and\
  \citenamefont {Ma}}]{2014LI}%
  \BibitemOpen
  \bibfield  {author} {\bibinfo {author} {\bibfnamefont {Y.}~\bibnamefont
  {Li}}, \bibinfo {author} {\bibfnamefont {J.}~\bibnamefont {Hao}}, \bibinfo
  {author} {\bibfnamefont {H.}~\bibnamefont {Liu}}, \bibinfo {author}
  {\bibfnamefont {Y.}~\bibnamefont {Li}}, \ and\ \bibinfo {author}
  {\bibfnamefont {Y.}~\bibnamefont {Ma}},\ }\href {\doibase 10.1063/1.4874158}
  {\bibfield  {journal} {\bibinfo  {journal} {J. Chem. Phys}\ }\textbf
  {\bibinfo {volume} {140}},\ \bibinfo {pages} {174712} (\bibinfo {year}
  {2014})}\BibitemShut {NoStop}%
\bibitem [{\citenamefont {Somayazulu}\ \emph {et~al.}(2019)\citenamefont
  {Somayazulu}, \citenamefont {Ahart}, \citenamefont {Mishra}, \citenamefont
  {Geballe}, \citenamefont {Baldini}, \citenamefont {Meng}, \citenamefont
  {Struzhkin},\ and\ \citenamefont {Hemley}}]{2019SOM}%
  \BibitemOpen
  \bibfield  {author} {\bibinfo {author} {\bibfnamefont {M.}~\bibnamefont
  {Somayazulu}}, \bibinfo {author} {\bibfnamefont {M.}~\bibnamefont {Ahart}},
  \bibinfo {author} {\bibfnamefont {A.~K.}\ \bibnamefont {Mishra}}, \bibinfo
  {author} {\bibfnamefont {Z.~M.}\ \bibnamefont {Geballe}}, \bibinfo {author}
  {\bibfnamefont {M.}~\bibnamefont {Baldini}}, \bibinfo {author} {\bibfnamefont
  {Y.}~\bibnamefont {Meng}}, \bibinfo {author} {\bibfnamefont {V.~V.}\
  \bibnamefont {Struzhkin}}, \ and\ \bibinfo {author} {\bibfnamefont {R.~J.}\
  \bibnamefont {Hemley}},\ }\href {\doibase 10.1103/physrevlett.122.027001}
  {\bibfield  {journal} {\bibinfo  {journal} {Phys. Rev. Lett.}\ }\textbf
  {\bibinfo {volume} {122}},\ \bibinfo {pages} {027001} (\bibinfo {year}
  {2019})}\BibitemShut {NoStop}%
\bibitem [{\citenamefont {Song}\ \emph {et~al.}(2021)\citenamefont {Song},
  \citenamefont {Hou}, \citenamefont {de~Castro}, \citenamefont {Nakano},
  \citenamefont {Hongo}, \citenamefont {Takano},\ and\ \citenamefont
  {Maezono}}]{2021SONa}%
  \BibitemOpen
  \bibfield  {author} {\bibinfo {author} {\bibfnamefont {P.}~\bibnamefont
  {Song}}, \bibinfo {author} {\bibfnamefont {Z.}~\bibnamefont {Hou}}, \bibinfo
  {author} {\bibfnamefont {P.~B.}\ \bibnamefont {de~Castro}}, \bibinfo {author}
  {\bibfnamefont {K.}~\bibnamefont {Nakano}}, \bibinfo {author} {\bibfnamefont
  {K.}~\bibnamefont {Hongo}}, \bibinfo {author} {\bibfnamefont
  {Y.}~\bibnamefont {Takano}}, \ and\ \bibinfo {author} {\bibfnamefont
  {R.}~\bibnamefont {Maezono}},\ }\href {\doibase
  10.1021/acs.chemmater.1c02371} {\bibfield  {journal} {\bibinfo  {journal}
  {Chem. Mater.}\ }\textbf {\bibinfo {volume} {33}},\ \bibinfo {pages} {9501}
  (\bibinfo {year} {2021})}\BibitemShut {NoStop}%
\bibitem [{\citenamefont {Ma}\ \emph {et~al.}(2017{\natexlab{a}})\citenamefont
  {Ma}, \citenamefont {Duan}, \citenamefont {Shao}, \citenamefont {Li},
  \citenamefont {Wang}, \citenamefont {Yu}, \citenamefont {Tian}, \citenamefont
  {Xie}, \citenamefont {Liu},\ and\ \citenamefont {Cui}}]{2017MAa}%
  \BibitemOpen
  \bibfield  {author} {\bibinfo {author} {\bibfnamefont {Y.}~\bibnamefont
  {Ma}}, \bibinfo {author} {\bibfnamefont {D.}~\bibnamefont {Duan}}, \bibinfo
  {author} {\bibfnamefont {Z.}~\bibnamefont {Shao}}, \bibinfo {author}
  {\bibfnamefont {D.}~\bibnamefont {Li}}, \bibinfo {author} {\bibfnamefont
  {L.}~\bibnamefont {Wang}}, \bibinfo {author} {\bibfnamefont {H.}~\bibnamefont
  {Yu}}, \bibinfo {author} {\bibfnamefont {F.}~\bibnamefont {Tian}}, \bibinfo
  {author} {\bibfnamefont {H.}~\bibnamefont {Xie}}, \bibinfo {author}
  {\bibfnamefont {B.}~\bibnamefont {Liu}}, \ and\ \bibinfo {author}
  {\bibfnamefont {T.}~\bibnamefont {Cui}},\ }\href {\doibase
  10.1039/c7cp05267g} {\bibfield  {journal} {\bibinfo  {journal} {Phys. Chem.
  Chem. Phys.}\ }\textbf {\bibinfo {volume} {19}},\ \bibinfo {pages} {27406}
  (\bibinfo {year} {2017}{\natexlab{a}})}\BibitemShut {NoStop}%
\bibitem [{\citenamefont {Ma}\ \emph {et~al.}(2017{\natexlab{b}})\citenamefont
  {Ma}, \citenamefont {Duan}, \citenamefont {Shao}, \citenamefont {Yu},
  \citenamefont {Liu}, \citenamefont {Tian}, \citenamefont {Huang},
  \citenamefont {Li}, \citenamefont {Liu},\ and\ \citenamefont
  {Cui}}]{2017MAb}%
  \BibitemOpen
  \bibfield  {author} {\bibinfo {author} {\bibfnamefont {Y.}~\bibnamefont
  {Ma}}, \bibinfo {author} {\bibfnamefont {D.}~\bibnamefont {Duan}}, \bibinfo
  {author} {\bibfnamefont {Z.}~\bibnamefont {Shao}}, \bibinfo {author}
  {\bibfnamefont {H.}~\bibnamefont {Yu}}, \bibinfo {author} {\bibfnamefont
  {H.}~\bibnamefont {Liu}}, \bibinfo {author} {\bibfnamefont {F.}~\bibnamefont
  {Tian}}, \bibinfo {author} {\bibfnamefont {X.}~\bibnamefont {Huang}},
  \bibinfo {author} {\bibfnamefont {D.}~\bibnamefont {Li}}, \bibinfo {author}
  {\bibfnamefont {B.}~\bibnamefont {Liu}}, \ and\ \bibinfo {author}
  {\bibfnamefont {T.}~\bibnamefont {Cui}},\ }\href {\doibase
  10.1103/physrevb.96.144518} {\bibfield  {journal} {\bibinfo  {journal} {Phys.
  Rev. B}\ }\textbf {\bibinfo {volume} {96}},\ \bibinfo {pages} {144518}
  (\bibinfo {year} {2017}{\natexlab{b}})}\BibitemShut {NoStop}%
\bibitem [{\citenamefont {Shao}\ \emph {et~al.}(2019)\citenamefont {Shao},
  \citenamefont {Duan}, \citenamefont {Ma}, \citenamefont {Yu}, \citenamefont
  {Song}, \citenamefont {Xie}, \citenamefont {Li}, \citenamefont {Tian},
  \citenamefont {Liu},\ and\ \citenamefont {Cui}}]{2019SHA}%
  \BibitemOpen
  \bibfield  {author} {\bibinfo {author} {\bibfnamefont {Z.}~\bibnamefont
  {Shao}}, \bibinfo {author} {\bibfnamefont {D.}~\bibnamefont {Duan}}, \bibinfo
  {author} {\bibfnamefont {Y.}~\bibnamefont {Ma}}, \bibinfo {author}
  {\bibfnamefont {H.}~\bibnamefont {Yu}}, \bibinfo {author} {\bibfnamefont
  {H.}~\bibnamefont {Song}}, \bibinfo {author} {\bibfnamefont {H.}~\bibnamefont
  {Xie}}, \bibinfo {author} {\bibfnamefont {D.}~\bibnamefont {Li}}, \bibinfo
  {author} {\bibfnamefont {F.}~\bibnamefont {Tian}}, \bibinfo {author}
  {\bibfnamefont {B.}~\bibnamefont {Liu}}, \ and\ \bibinfo {author}
  {\bibfnamefont {T.}~\bibnamefont {Cui}},\ }\href {\doibase
  10.1038/s41524-019-0244-6} {\bibfield  {journal} {\bibinfo  {journal} {npj
  Comput. Mater.}\ }\textbf {\bibinfo {volume} {5}},\ \bibinfo {pages} {1}
  (\bibinfo {year} {2019})}\BibitemShut {NoStop}%
\bibitem [{\citenamefont {Zheng}\ \emph {et~al.}(2021)\citenamefont {Zheng},
  \citenamefont {Sun}, \citenamefont {Dou}, \citenamefont {Mao},\ and\
  \citenamefont {Lu}}]{2021ZHE}%
  \BibitemOpen
  \bibfield  {author} {\bibinfo {author} {\bibfnamefont {J.}~\bibnamefont
  {Zheng}}, \bibinfo {author} {\bibfnamefont {W.}~\bibnamefont {Sun}}, \bibinfo
  {author} {\bibfnamefont {X.}~\bibnamefont {Dou}}, \bibinfo {author}
  {\bibfnamefont {A.-J.}\ \bibnamefont {Mao}}, \ and\ \bibinfo {author}
  {\bibfnamefont {C.}~\bibnamefont {Lu}},\ }\href {\doibase
  10.1021/acs.jpcc.0c09447} {\bibfield  {journal} {\bibinfo  {journal} {J.
  Phys. Chem. C}\ }\textbf {\bibinfo {volume} {125}},\ \bibinfo {pages} {3150}
  (\bibinfo {year} {2021})}\BibitemShut {NoStop}%
\bibitem [{\citenamefont {Rahm}\ \emph {et~al.}(2017)\citenamefont {Rahm},
  \citenamefont {Hoffmann},\ and\ \citenamefont {Ashcroft}}]{2017RAH}%
  \BibitemOpen
  \bibfield  {author} {\bibinfo {author} {\bibfnamefont {M.}~\bibnamefont
  {Rahm}}, \bibinfo {author} {\bibfnamefont {R.}~\bibnamefont {Hoffmann}}, \
  and\ \bibinfo {author} {\bibfnamefont {N.~W.}\ \bibnamefont {Ashcroft}},\
  }\href {\doibase 10.1021/jacs.7b04456} {\bibfield  {journal} {\bibinfo
  {journal} {J. Am. Chem. Soc.}\ }\textbf {\bibinfo {volume} {139}},\ \bibinfo
  {pages} {8740} (\bibinfo {year} {2017})}\BibitemShut {NoStop}%
\bibitem [{\citenamefont {Liang}\ \emph {et~al.}(2021)\citenamefont {Liang},
  \citenamefont {Bergara}, \citenamefont {Wei}, \citenamefont {Song},
  \citenamefont {Wang}, \citenamefont {Sun}, \citenamefont {Liu}, \citenamefont
  {Hemley}, \citenamefont {Wang}, \citenamefont {Gao},\ and\ \citenamefont
  {Tian}}]{2021LIA}%
  \BibitemOpen
  \bibfield  {author} {\bibinfo {author} {\bibfnamefont {X.}~\bibnamefont
  {Liang}}, \bibinfo {author} {\bibfnamefont {A.}~\bibnamefont {Bergara}},
  \bibinfo {author} {\bibfnamefont {X.}~\bibnamefont {Wei}}, \bibinfo {author}
  {\bibfnamefont {X.}~\bibnamefont {Song}}, \bibinfo {author} {\bibfnamefont
  {L.}~\bibnamefont {Wang}}, \bibinfo {author} {\bibfnamefont {R.}~\bibnamefont
  {Sun}}, \bibinfo {author} {\bibfnamefont {H.}~\bibnamefont {Liu}}, \bibinfo
  {author} {\bibfnamefont {R.~J.}\ \bibnamefont {Hemley}}, \bibinfo {author}
  {\bibfnamefont {L.}~\bibnamefont {Wang}}, \bibinfo {author} {\bibfnamefont
  {G.}~\bibnamefont {Gao}}, \ and\ \bibinfo {author} {\bibfnamefont
  {Y.}~\bibnamefont {Tian}},\ }\href {\doibase 10.1103/physrevb.104.134501}
  {\bibfield  {journal} {\bibinfo  {journal} {Phys. Rev. B}\ }\textbf {\bibinfo
  {volume} {104}},\ \bibinfo {pages} {134501} (\bibinfo {year}
  {2021})}\BibitemShut {NoStop}%
\bibitem [{\citenamefont {Sun}\ \emph {et~al.}(2019)\citenamefont {Sun},
  \citenamefont {Lv}, \citenamefont {Xie}, \citenamefont {Liu},\ and\
  \citenamefont {Ma}}]{2019SUN}%
  \BibitemOpen
  \bibfield  {author} {\bibinfo {author} {\bibfnamefont {Y.}~\bibnamefont
  {Sun}}, \bibinfo {author} {\bibfnamefont {J.}~\bibnamefont {Lv}}, \bibinfo
  {author} {\bibfnamefont {Y.}~\bibnamefont {Xie}}, \bibinfo {author}
  {\bibfnamefont {H.}~\bibnamefont {Liu}}, \ and\ \bibinfo {author}
  {\bibfnamefont {Y.}~\bibnamefont {Ma}},\ }\href {\doibase
  10.1103/physrevlett.123.097001} {\bibfield  {journal} {\bibinfo  {journal}
  {Phys. Rev. Lett.}\ }\textbf {\bibinfo {volume} {123}},\ \bibinfo {pages}
  {097001} (\bibinfo {year} {2019})}\BibitemShut {NoStop}%
\bibitem [{\citenamefont {Liang}\ \emph
  {et~al.}(2019{\natexlab{a}})\citenamefont {Liang}, \citenamefont {Zhao},
  \citenamefont {Shao}, \citenamefont {Bergara}, \citenamefont {Liu},
  \citenamefont {Wang}, \citenamefont {Sun}, \citenamefont {Zhang},
  \citenamefont {Gao}, \citenamefont {Zhao}, \citenamefont {Zhou},
  \citenamefont {He}, \citenamefont {Yu}, \citenamefont {Gao},\ and\
  \citenamefont {Tian}}]{2019LIAa}%
  \BibitemOpen
  \bibfield  {author} {\bibinfo {author} {\bibfnamefont {X.}~\bibnamefont
  {Liang}}, \bibinfo {author} {\bibfnamefont {S.}~\bibnamefont {Zhao}},
  \bibinfo {author} {\bibfnamefont {C.}~\bibnamefont {Shao}}, \bibinfo {author}
  {\bibfnamefont {A.}~\bibnamefont {Bergara}}, \bibinfo {author} {\bibfnamefont
  {H.}~\bibnamefont {Liu}}, \bibinfo {author} {\bibfnamefont {L.}~\bibnamefont
  {Wang}}, \bibinfo {author} {\bibfnamefont {R.}~\bibnamefont {Sun}}, \bibinfo
  {author} {\bibfnamefont {Y.}~\bibnamefont {Zhang}}, \bibinfo {author}
  {\bibfnamefont {Y.}~\bibnamefont {Gao}}, \bibinfo {author} {\bibfnamefont
  {Z.}~\bibnamefont {Zhao}}, \bibinfo {author} {\bibfnamefont {X.-F.}\
  \bibnamefont {Zhou}}, \bibinfo {author} {\bibfnamefont {J.}~\bibnamefont
  {He}}, \bibinfo {author} {\bibfnamefont {D.}~\bibnamefont {Yu}}, \bibinfo
  {author} {\bibfnamefont {G.}~\bibnamefont {Gao}}, \ and\ \bibinfo {author}
  {\bibfnamefont {Y.}~\bibnamefont {Tian}},\ }\href {\doibase
  10.1103/physrevb.100.184502} {\bibfield  {journal} {\bibinfo  {journal}
  {Phys. Rev. B}\ }\textbf {\bibinfo {volume} {100}},\ \bibinfo {pages}
  {184502} (\bibinfo {year} {2019}{\natexlab{a}})}\BibitemShut {NoStop}%
\bibitem [{\citenamefont {Wei}\ \emph {et~al.}(2020)\citenamefont {Wei},
  \citenamefont {Jia}, \citenamefont {Fang}, \citenamefont {Wang},
  \citenamefont {Qian}, \citenamefont {Yuan}, \citenamefont {Selvaraj},
  \citenamefont {Ji},\ and\ \citenamefont {Wei}}]{2020WEI}%
  \BibitemOpen
  \bibfield  {author} {\bibinfo {author} {\bibfnamefont {Y.~K.}\ \bibnamefont
  {Wei}}, \bibinfo {author} {\bibfnamefont {L.~Q.}\ \bibnamefont {Jia}},
  \bibinfo {author} {\bibfnamefont {Y.~Y.}\ \bibnamefont {Fang}}, \bibinfo
  {author} {\bibfnamefont {L.~J.}\ \bibnamefont {Wang}}, \bibinfo {author}
  {\bibfnamefont {Z.~X.}\ \bibnamefont {Qian}}, \bibinfo {author}
  {\bibfnamefont {J.~N.}\ \bibnamefont {Yuan}}, \bibinfo {author}
  {\bibfnamefont {G.}~\bibnamefont {Selvaraj}}, \bibinfo {author}
  {\bibfnamefont {G.~F.}\ \bibnamefont {Ji}}, \ and\ \bibinfo {author}
  {\bibfnamefont {D.~Q.}\ \bibnamefont {Wei}},\ }\href {\doibase
  10.1002/qua.26459} {\bibfield  {journal} {\bibinfo  {journal} {Int. J.
  Quantum Chem.}\ }\textbf {\bibinfo {volume} {121}},\ \bibinfo {pages} {26459}
  (\bibinfo {year} {2020})}\BibitemShut {NoStop}%
\bibitem [{\citenamefont {Shi}\ \emph {et~al.}(2021)\citenamefont {Shi},
  \citenamefont {Wei}, \citenamefont {Liang}, \citenamefont {Turnbull},
  \citenamefont {Cheng}, \citenamefont {Chen},\ and\ \citenamefont
  {Ji}}]{2021SHI}%
  \BibitemOpen
  \bibfield  {author} {\bibinfo {author} {\bibfnamefont {L.-T.}\ \bibnamefont
  {Shi}}, \bibinfo {author} {\bibfnamefont {Y.-K.}\ \bibnamefont {Wei}},
  \bibinfo {author} {\bibfnamefont {A.-K.}\ \bibnamefont {Liang}}, \bibinfo
  {author} {\bibfnamefont {R.}~\bibnamefont {Turnbull}}, \bibinfo {author}
  {\bibfnamefont {C.}~\bibnamefont {Cheng}}, \bibinfo {author} {\bibfnamefont
  {X.-R.}\ \bibnamefont {Chen}}, \ and\ \bibinfo {author} {\bibfnamefont
  {G.-F.}\ \bibnamefont {Ji}},\ }\href {\doibase 10.1039/d1tc00634g} {\bibfield
   {journal} {\bibinfo  {journal} {J. Mater. Chem. C}\ }\textbf {\bibinfo
  {volume} {9}},\ \bibinfo {pages} {7284} (\bibinfo {year} {2021})}\BibitemShut
  {NoStop}%
\bibitem [{\citenamefont {Song}\ \emph
  {et~al.}(2022{\natexlab{a}})\citenamefont {Song}, \citenamefont {Hou},
  \citenamefont {de~Castro}, \citenamefont {Nakano}, \citenamefont {Takano},
  \citenamefont {Maezono},\ and\ \citenamefont {Hongo}}]{2021SONb}%
  \BibitemOpen
  \bibfield  {author} {\bibinfo {author} {\bibfnamefont {P.}~\bibnamefont
  {Song}}, \bibinfo {author} {\bibfnamefont {Z.}~\bibnamefont {Hou}}, \bibinfo
  {author} {\bibfnamefont {P.~B.}\ \bibnamefont {de~Castro}}, \bibinfo {author}
  {\bibfnamefont {K.}~\bibnamefont {Nakano}}, \bibinfo {author} {\bibfnamefont
  {Y.}~\bibnamefont {Takano}}, \bibinfo {author} {\bibfnamefont
  {R.}~\bibnamefont {Maezono}}, \ and\ \bibinfo {author} {\bibfnamefont
  {K.}~\bibnamefont {Hongo}},\ }\href {\doibase 10.1002/adts.202100364}
  {\bibfield  {journal} {\bibinfo  {journal} {Adv. Theory Simul.}\ }\textbf
  {\bibinfo {volume} {5}},\ \bibinfo {pages} {2100364} (\bibinfo {year}
  {2022}{\natexlab{a}})}\BibitemShut {NoStop}%
\bibitem [{\citenamefont {Semenok}\ \emph {et~al.}(2021)\citenamefont
  {Semenok}, \citenamefont {Troyan}, \citenamefont {Ivanova}, \citenamefont
  {Kvashnin}, \citenamefont {Kruglov}, \citenamefont {Hanfland}, \citenamefont
  {Sadakov}, \citenamefont {Sobolevskiy}, \citenamefont {Pervakov},
  \citenamefont {Lyubutin}, \citenamefont {Glazyrin}, \citenamefont {Giordano},
  \citenamefont {Karimov}, \citenamefont {Vasiliev}, \citenamefont {Akashi},
  \citenamefont {Pudalov},\ and\ \citenamefont {Oganov}}]{2021SEM}%
  \BibitemOpen
  \bibfield  {author} {\bibinfo {author} {\bibfnamefont {D.~V.}\ \bibnamefont
  {Semenok}}, \bibinfo {author} {\bibfnamefont {I.~A.}\ \bibnamefont {Troyan}},
  \bibinfo {author} {\bibfnamefont {A.~G.}\ \bibnamefont {Ivanova}}, \bibinfo
  {author} {\bibfnamefont {A.~G.}\ \bibnamefont {Kvashnin}}, \bibinfo {author}
  {\bibfnamefont {I.~A.}\ \bibnamefont {Kruglov}}, \bibinfo {author}
  {\bibfnamefont {M.}~\bibnamefont {Hanfland}}, \bibinfo {author}
  {\bibfnamefont {A.~V.}\ \bibnamefont {Sadakov}}, \bibinfo {author}
  {\bibfnamefont {O.~A.}\ \bibnamefont {Sobolevskiy}}, \bibinfo {author}
  {\bibfnamefont {K.~S.}\ \bibnamefont {Pervakov}}, \bibinfo {author}
  {\bibfnamefont {I.~S.}\ \bibnamefont {Lyubutin}}, \bibinfo {author}
  {\bibfnamefont {K.~V.}\ \bibnamefont {Glazyrin}}, \bibinfo {author}
  {\bibfnamefont {N.}~\bibnamefont {Giordano}}, \bibinfo {author}
  {\bibfnamefont {D.~N.}\ \bibnamefont {Karimov}}, \bibinfo {author}
  {\bibfnamefont {A.~L.}\ \bibnamefont {Vasiliev}}, \bibinfo {author}
  {\bibfnamefont {R.}~\bibnamefont {Akashi}}, \bibinfo {author} {\bibfnamefont
  {V.~M.}\ \bibnamefont {Pudalov}}, \ and\ \bibinfo {author} {\bibfnamefont
  {A.~R.}\ \bibnamefont {Oganov}},\ }\href {\doibase
  10.1016/j.mattod.2021.03.025} {\bibfield  {journal} {\bibinfo  {journal}
  {Mater. Today}\ }\textbf {\bibinfo {volume} {48}},\ \bibinfo {pages} {18}
  (\bibinfo {year} {2021})}\BibitemShut {NoStop}%
\bibitem [{\citenamefont {Liang}\ \emph
  {et~al.}(2019{\natexlab{b}})\citenamefont {Liang}, \citenamefont {Bergara},
  \citenamefont {Wang}, \citenamefont {Wen}, \citenamefont {Zhao},
  \citenamefont {Zhou}, \citenamefont {He}, \citenamefont {Gao},\ and\
  \citenamefont {Tian}}]{2019LIAb}%
  \BibitemOpen
  \bibfield  {author} {\bibinfo {author} {\bibfnamefont {X.}~\bibnamefont
  {Liang}}, \bibinfo {author} {\bibfnamefont {A.}~\bibnamefont {Bergara}},
  \bibinfo {author} {\bibfnamefont {L.}~\bibnamefont {Wang}}, \bibinfo {author}
  {\bibfnamefont {B.}~\bibnamefont {Wen}}, \bibinfo {author} {\bibfnamefont
  {Z.}~\bibnamefont {Zhao}}, \bibinfo {author} {\bibfnamefont {X.-F.}\
  \bibnamefont {Zhou}}, \bibinfo {author} {\bibfnamefont {J.}~\bibnamefont
  {He}}, \bibinfo {author} {\bibfnamefont {G.}~\bibnamefont {Gao}}, \ and\
  \bibinfo {author} {\bibfnamefont {Y.}~\bibnamefont {Tian}},\ }\href {\doibase
  10.1103/physrevb.99.100505} {\bibfield  {journal} {\bibinfo  {journal} {Phys.
  Rev. B}\ }\textbf {\bibinfo {volume} {99}},\ \bibinfo {pages} {100505}
  (\bibinfo {year} {2019}{\natexlab{b}})}\BibitemShut {NoStop}%
\bibitem [{\citenamefont {Jiang}\ \emph {et~al.}(2021)\citenamefont {Jiang},
  \citenamefont {Tian}, \citenamefont {Hai}, \citenamefont {Lu}, \citenamefont
  {Tong}, \citenamefont {Wu}, \citenamefont {Li}, \citenamefont {Yang},\ and\
  \citenamefont {Zhong}}]{2021JIA}%
  \BibitemOpen
  \bibfield  {author} {\bibinfo {author} {\bibfnamefont {M.-J.}\ \bibnamefont
  {Jiang}}, \bibinfo {author} {\bibfnamefont {H.-L.}\ \bibnamefont {Tian}},
  \bibinfo {author} {\bibfnamefont {Y.-L.}\ \bibnamefont {Hai}}, \bibinfo
  {author} {\bibfnamefont {N.}~\bibnamefont {Lu}}, \bibinfo {author}
  {\bibfnamefont {P.-F.}\ \bibnamefont {Tong}}, \bibinfo {author}
  {\bibfnamefont {S.-Y.}\ \bibnamefont {Wu}}, \bibinfo {author} {\bibfnamefont
  {W.-J.}\ \bibnamefont {Li}}, \bibinfo {author} {\bibfnamefont {C.-L.}\
  \bibnamefont {Yang}}, \ and\ \bibinfo {author} {\bibfnamefont {G.-H.}\
  \bibnamefont {Zhong}},\ }\href {\doibase 10.1021/acsaelm.1c00617} {\bibfield
  {journal} {\bibinfo  {journal} {ACS Appl. Electron. Mater}\ }\textbf
  {\bibinfo {volume} {3}},\ \bibinfo {pages} {4172} (\bibinfo {year}
  {2021})}\BibitemShut {NoStop}%
\bibitem [{\citenamefont {Cui}\ \emph {et~al.}(2020)\citenamefont {Cui},
  \citenamefont {Bi}, \citenamefont {Shi}, \citenamefont {Li}, \citenamefont
  {Liu}, \citenamefont {Zurek},\ and\ \citenamefont {Hemley}}]{2020CUI}%
  \BibitemOpen
  \bibfield  {author} {\bibinfo {author} {\bibfnamefont {W.}~\bibnamefont
  {Cui}}, \bibinfo {author} {\bibfnamefont {T.}~\bibnamefont {Bi}}, \bibinfo
  {author} {\bibfnamefont {J.}~\bibnamefont {Shi}}, \bibinfo {author}
  {\bibfnamefont {Y.}~\bibnamefont {Li}}, \bibinfo {author} {\bibfnamefont
  {H.}~\bibnamefont {Liu}}, \bibinfo {author} {\bibfnamefont {E.}~\bibnamefont
  {Zurek}}, \ and\ \bibinfo {author} {\bibfnamefont {R.~J.}\ \bibnamefont
  {Hemley}},\ }\href {\doibase 10.1103/physrevb.101.134504} {\bibfield
  {journal} {\bibinfo  {journal} {Phys. Rev. B}\ }\textbf {\bibinfo {volume}
  {101}},\ \bibinfo {pages} {134504} (\bibinfo {year} {2020})}\BibitemShut
  {NoStop}%
\bibitem [{\citenamefont {Cataldo}\ \emph {et~al.}(2020)\citenamefont
  {Cataldo}, \citenamefont {von~der Linden},\ and\ \citenamefont
  {Boeri}}]{2020CAT}%
  \BibitemOpen
  \bibfield  {author} {\bibinfo {author} {\bibfnamefont {S.~D.}\ \bibnamefont
  {Cataldo}}, \bibinfo {author} {\bibfnamefont {W.}~\bibnamefont {von~der
  Linden}}, \ and\ \bibinfo {author} {\bibfnamefont {L.}~\bibnamefont
  {Boeri}},\ }\href {\doibase 10.1103/physrevb.102.014516} {\bibfield
  {journal} {\bibinfo  {journal} {Phys. Rev. B}\ }\textbf {\bibinfo {volume}
  {102}},\ \bibinfo {pages} {014516} (\bibinfo {year} {2020})}\BibitemShut
  {NoStop}%
\bibitem [{\citenamefont {Song}\ \emph
  {et~al.}(2022{\natexlab{b}})\citenamefont {Song}, \citenamefont {Hou},
  \citenamefont {de~Castro}, \citenamefont {Nakano}, \citenamefont {Hongo},
  \citenamefont {Takano},\ and\ \citenamefont {Maezono}}]{2021SONc}%
  \BibitemOpen
  \bibfield  {author} {\bibinfo {author} {\bibfnamefont {P.}~\bibnamefont
  {Song}}, \bibinfo {author} {\bibfnamefont {Z.}~\bibnamefont {Hou}}, \bibinfo
  {author} {\bibfnamefont {P.~B.}\ \bibnamefont {de~Castro}}, \bibinfo {author}
  {\bibfnamefont {K.}~\bibnamefont {Nakano}}, \bibinfo {author} {\bibfnamefont
  {K.}~\bibnamefont {Hongo}}, \bibinfo {author} {\bibfnamefont
  {Y.}~\bibnamefont {Takano}}, \ and\ \bibinfo {author} {\bibfnamefont
  {R.}~\bibnamefont {Maezono}},\ }\href {\doibase 10.1021/acs.jpcc.1c08743}
  {\bibfield  {journal} {\bibinfo  {journal} {J. Phys. Chem. C}\ }\textbf
  {\bibinfo {volume} {126}},\ \bibinfo {pages} {2747} (\bibinfo {year}
  {2022}{\natexlab{b}})}\BibitemShut {NoStop}%
\bibitem [{\citenamefont {Cataldo}\ \emph {et~al.}(2022)\citenamefont
  {Cataldo}, \citenamefont {von~der Linden},\ and\ \citenamefont
  {Boeri}}]{2022CAT}%
  \BibitemOpen
  \bibfield  {author} {\bibinfo {author} {\bibfnamefont {S.~D.}\ \bibnamefont
  {Cataldo}}, \bibinfo {author} {\bibfnamefont {W.}~\bibnamefont {von~der
  Linden}}, \ and\ \bibinfo {author} {\bibfnamefont {L.}~\bibnamefont
  {Boeri}},\ }\href {\doibase 10.1038/s41524-021-00691-6} {\bibfield  {journal}
  {\bibinfo  {journal} {Npj Comput. Mater.}\ }\textbf {\bibinfo {volume} {8}},\
  \bibinfo {pages} {1} (\bibinfo {year} {2022})}\BibitemShut {NoStop}%
\bibitem [{\citenamefont {Vocaturo}\ \emph {et~al.}(2022)\citenamefont
  {Vocaturo}, \citenamefont {Tresca}, \citenamefont {Ghiringhelli},\ and\
  \citenamefont {Profeta}}]{2022VOC}%
  \BibitemOpen
  \bibfield  {author} {\bibinfo {author} {\bibfnamefont {R.}~\bibnamefont
  {Vocaturo}}, \bibinfo {author} {\bibfnamefont {C.}~\bibnamefont {Tresca}},
  \bibinfo {author} {\bibfnamefont {G.}~\bibnamefont {Ghiringhelli}}, \ and\
  \bibinfo {author} {\bibfnamefont {G.}~\bibnamefont {Profeta}},\ }\href
  {\doibase 10.1063/5.0076728} {\bibfield  {journal} {\bibinfo  {journal} {J.
  Appl. Phys.}\ }\textbf {\bibinfo {volume} {131}},\ \bibinfo {pages} {033903}
  (\bibinfo {year} {2022})}\BibitemShut {NoStop}%
\bibitem [{\citenamefont {Snider}\ \emph {et~al.}(2020)\citenamefont {Snider},
  \citenamefont {Dasenbrock-Gammon}, \citenamefont {McBride}, \citenamefont
  {Debessai}, \citenamefont {Vindana}, \citenamefont {Vencatasamy},
  \citenamefont {Lawler}, \citenamefont {Salamat},\ and\ \citenamefont
  {Dias}}]{2020SNI}%
  \BibitemOpen
  \bibfield  {author} {\bibinfo {author} {\bibfnamefont {E.}~\bibnamefont
  {Snider}}, \bibinfo {author} {\bibfnamefont {N.}~\bibnamefont
  {Dasenbrock-Gammon}}, \bibinfo {author} {\bibfnamefont {R.}~\bibnamefont
  {McBride}}, \bibinfo {author} {\bibfnamefont {M.}~\bibnamefont {Debessai}},
  \bibinfo {author} {\bibfnamefont {H.}~\bibnamefont {Vindana}}, \bibinfo
  {author} {\bibfnamefont {K.}~\bibnamefont {Vencatasamy}}, \bibinfo {author}
  {\bibfnamefont {K.~V.}\ \bibnamefont {Lawler}}, \bibinfo {author}
  {\bibfnamefont {A.}~\bibnamefont {Salamat}}, \ and\ \bibinfo {author}
  {\bibfnamefont {R.~P.}\ \bibnamefont {Dias}},\ }\href {\doibase
  10.1038/s41586-020-2801-z} {\bibfield  {journal} {\bibinfo  {journal}
  {Nature}\ }\textbf {\bibinfo {volume} {586}},\ \bibinfo {pages} {373}
  (\bibinfo {year} {2020})}\BibitemShut {NoStop}%
\bibitem [{\citenamefont {Hirsch}\ and\ \citenamefont
  {Marsiglio}(2021)}]{2021HIR}%
  \BibitemOpen
  \bibfield  {author} {\bibinfo {author} {\bibfnamefont {J.~E.}\ \bibnamefont
  {Hirsch}}\ and\ \bibinfo {author} {\bibfnamefont {F.}~\bibnamefont
  {Marsiglio}},\ }\href {\doibase 10.1038/s41586-021-03595-z} {\bibfield
  {journal} {\bibinfo  {journal} {Nature}\ }\textbf {\bibinfo {volume} {596}},\
  \bibinfo {pages} {E9} (\bibinfo {year} {2021})}\BibitemShut {NoStop}%
\bibitem [{\citenamefont {Hirsch}(2022)}]{2022HIR}%
  \BibitemOpen
  \bibfield  {author} {\bibinfo {author} {\bibfnamefont {J.~E.}\ \bibnamefont
  {Hirsch}},\ }\href {\doibase 10.1209/0295-5075/ac50c9} {\bibfield  {journal}
  {\bibinfo  {journal} {Europhys. Lett.}\ }\textbf {\bibinfo {volume} {137}},\
  \bibinfo {pages} {36001} (\bibinfo {year} {2022})}\BibitemShut {NoStop}%
\bibitem [{\citenamefont {Nakanishi}\ \emph {et~al.}(2018)\citenamefont
  {Nakanishi}, \citenamefont {Ishikawa},\ and\ \citenamefont
  {Shimizu}}]{2018NAK}%
  \BibitemOpen
  \bibfield  {author} {\bibinfo {author} {\bibfnamefont {A.}~\bibnamefont
  {Nakanishi}}, \bibinfo {author} {\bibfnamefont {T.}~\bibnamefont {Ishikawa}},
  \ and\ \bibinfo {author} {\bibfnamefont {K.}~\bibnamefont {Shimizu}},\ }\href
  {\doibase 10.7566/jpsj.87.124711} {\bibfield  {journal} {\bibinfo  {journal}
  {J. Phys. Soc. Jpn.}\ }\textbf {\bibinfo {volume} {87}},\ \bibinfo {pages}
  {124711} (\bibinfo {year} {2018})}\BibitemShut {NoStop}%
\bibitem [{\citenamefont {Amsler}(2019)}]{2019AMS}%
  \BibitemOpen
  \bibfield  {author} {\bibinfo {author} {\bibfnamefont {M.}~\bibnamefont
  {Amsler}},\ }\href {\doibase 10.1103/physrevb.99.060102} {\bibfield
  {journal} {\bibinfo  {journal} {Phys. Rev. B}\ }\textbf {\bibinfo {volume}
  {99}},\ \bibinfo {pages} {060102} (\bibinfo {year} {2019})}\BibitemShut
  {NoStop}%
\bibitem [{\citenamefont {Guan}\ \emph {et~al.}(2021)\citenamefont {Guan},
  \citenamefont {Sun},\ and\ \citenamefont {Liu}}]{2021GUA}%
  \BibitemOpen
  \bibfield  {author} {\bibinfo {author} {\bibfnamefont {H.}~\bibnamefont
  {Guan}}, \bibinfo {author} {\bibfnamefont {Y.}~\bibnamefont {Sun}}, \ and\
  \bibinfo {author} {\bibfnamefont {H.}~\bibnamefont {Liu}},\ }\href {\doibase
  10.1103/physrevresearch.3.043102} {\bibfield  {journal} {\bibinfo  {journal}
  {Phys. Rev. Res.}\ }\textbf {\bibinfo {volume} {3}},\ \bibinfo {pages}
  {043102} (\bibinfo {year} {2021})}\BibitemShut {NoStop}%
\bibitem [{\citenamefont {Chen}\ \emph
  {et~al.}(2022{\natexlab{a}})\citenamefont {Chen}, \citenamefont {Huang},
  \citenamefont {Semenok}, \citenamefont {Chen}, \citenamefont {Zhang},
  \citenamefont {Oganov},\ and\ \citenamefont {Cui}}]{2022CHEb}%
  \BibitemOpen
  \bibfield  {author} {\bibinfo {author} {\bibfnamefont {W.}~\bibnamefont
  {Chen}}, \bibinfo {author} {\bibfnamefont {X.}~\bibnamefont {Huang}},
  \bibinfo {author} {\bibfnamefont {D.~V.}\ \bibnamefont {Semenok}}, \bibinfo
  {author} {\bibfnamefont {S.}~\bibnamefont {Chen}}, \bibinfo {author}
  {\bibfnamefont {K.}~\bibnamefont {Zhang}}, \bibinfo {author} {\bibfnamefont
  {A.~R.}\ \bibnamefont {Oganov}}, \ and\ \bibinfo {author} {\bibfnamefont
  {T.}~\bibnamefont {Cui}},\ }\href@noop {} {\bibfield  {journal} {\bibinfo
  {journal} {arXiv preprint arXiv:2203.14353}\ } (\bibinfo {year}
  {2022}{\natexlab{a}})}\BibitemShut {NoStop}%
\bibitem [{\citenamefont {Glass}\ \emph {et~al.}(2006)\citenamefont {Glass},
  \citenamefont {Oganov},\ and\ \citenamefont {Hansen}}]{2006GLA}%
  \BibitemOpen
  \bibfield  {author} {\bibinfo {author} {\bibfnamefont {C.~W.}\ \bibnamefont
  {Glass}}, \bibinfo {author} {\bibfnamefont {A.~R.}\ \bibnamefont {Oganov}}, \
  and\ \bibinfo {author} {\bibfnamefont {N.}~\bibnamefont {Hansen}},\ }\href
  {\doibase 10.1016/j.cpc.2006.07.020} {\bibfield  {journal} {\bibinfo
  {journal} {Comput. Phys. Commun.}\ }\textbf {\bibinfo {volume} {175}},\
  \bibinfo {pages} {713} (\bibinfo {year} {2006})}\BibitemShut {NoStop}%
\bibitem [{\citenamefont {Lyakhov}\ \emph {et~al.}(2013)\citenamefont
  {Lyakhov}, \citenamefont {Oganov}, \citenamefont {Stokes},\ and\
  \citenamefont {Zhu}}]{2013LYA}%
  \BibitemOpen
  \bibfield  {author} {\bibinfo {author} {\bibfnamefont {A.~O.}\ \bibnamefont
  {Lyakhov}}, \bibinfo {author} {\bibfnamefont {A.~R.}\ \bibnamefont {Oganov}},
  \bibinfo {author} {\bibfnamefont {H.~T.}\ \bibnamefont {Stokes}}, \ and\
  \bibinfo {author} {\bibfnamefont {Q.}~\bibnamefont {Zhu}},\ }\href {\doibase
  10.1016/j.cpc.2012.12.009} {\bibfield  {journal} {\bibinfo  {journal}
  {Comput. Phys. Commun.}\ }\textbf {\bibinfo {volume} {184}},\ \bibinfo
  {pages} {1172} (\bibinfo {year} {2013})}\BibitemShut {NoStop}%
\bibitem [{\citenamefont {Perdew}\ \emph {et~al.}(1996)\citenamefont {Perdew},
  \citenamefont {Burke},\ and\ \citenamefont {Ernzerhof}}]{Perdew1996prl}%
  \BibitemOpen
  \bibfield  {author} {\bibinfo {author} {\bibfnamefont {J.~P.}\ \bibnamefont
  {Perdew}}, \bibinfo {author} {\bibfnamefont {K.}~\bibnamefont {Burke}}, \
  and\ \bibinfo {author} {\bibfnamefont {M.}~\bibnamefont {Ernzerhof}},\ }\href
  {\doibase 10.1103/PhysRevLett.77.3865} {\bibfield  {journal} {\bibinfo
  {journal} {Phys. Rev. Lett.}\ }\textbf {\bibinfo {volume} {77}},\ \bibinfo
  {pages} {3865} (\bibinfo {year} {1996})}\BibitemShut {NoStop}%
\bibitem [{\citenamefont {Kresse}\ and\ \citenamefont
  {Furthm\"uller}(1996)}]{Kresse1996prb}%
  \BibitemOpen
  \bibfield  {author} {\bibinfo {author} {\bibfnamefont {G.}~\bibnamefont
  {Kresse}}\ and\ \bibinfo {author} {\bibfnamefont {J.}~\bibnamefont
  {Furthm\"uller}},\ }\href {\doibase 10.1103/PhysRevB.54.11169} {\bibfield
  {journal} {\bibinfo  {journal} {Phys. Rev. B}\ }\textbf {\bibinfo {volume}
  {54}},\ \bibinfo {pages} {11169} (\bibinfo {year} {1996})}\BibitemShut
  {NoStop}%
\bibitem [{\citenamefont {Kresse}\ and\ \citenamefont
  {Joubert}(1999)}]{Kresse1999prb}%
  \BibitemOpen
  \bibfield  {author} {\bibinfo {author} {\bibfnamefont {G.}~\bibnamefont
  {Kresse}}\ and\ \bibinfo {author} {\bibfnamefont {D.}~\bibnamefont
  {Joubert}},\ }\href {\doibase 10.1103/PhysRevB.59.1758} {\bibfield  {journal}
  {\bibinfo  {journal} {Phys. Rev. B}\ }\textbf {\bibinfo {volume} {59}},\
  \bibinfo {pages} {1758} (\bibinfo {year} {1999})}\BibitemShut {NoStop}%
\bibitem [{\citenamefont {M{\"{u}}ller}\ \emph {et~al.}(2021)\citenamefont
  {M{\"{u}}ller}, \citenamefont {Ertural}, \citenamefont {Hempelmann},\ and\
  \citenamefont {Dronskowski}}]{2021MUL}%
  \BibitemOpen
  \bibfield  {author} {\bibinfo {author} {\bibfnamefont {P.~C.}\ \bibnamefont
  {M{\"{u}}ller}}, \bibinfo {author} {\bibfnamefont {C.}~\bibnamefont
  {Ertural}}, \bibinfo {author} {\bibfnamefont {J.}~\bibnamefont {Hempelmann}},
  \ and\ \bibinfo {author} {\bibfnamefont {R.}~\bibnamefont {Dronskowski}},\
  }\href {\doibase 10.1021/acs.jpcc.1c00718} {\bibfield  {journal} {\bibinfo
  {journal} {J. Phys. Chem. C.}\ }\textbf {\bibinfo {volume} {125}},\ \bibinfo
  {pages} {7959} (\bibinfo {year} {2021})}\BibitemShut {NoStop}%
\bibitem [{\citenamefont {Dudarev}\ \emph {et~al.}(1998)\citenamefont
  {Dudarev}, \citenamefont {Botton}, \citenamefont {Savrasov}, \citenamefont
  {Humphreys},\ and\ \citenamefont {Sutton}}]{1998DBS}%
  \BibitemOpen
  \bibfield  {author} {\bibinfo {author} {\bibfnamefont {S.~L.}\ \bibnamefont
  {Dudarev}}, \bibinfo {author} {\bibfnamefont {G.~A.}\ \bibnamefont {Botton}},
  \bibinfo {author} {\bibfnamefont {S.~Y.}\ \bibnamefont {Savrasov}}, \bibinfo
  {author} {\bibfnamefont {C.~J.}\ \bibnamefont {Humphreys}}, \ and\ \bibinfo
  {author} {\bibfnamefont {A.~P.}\ \bibnamefont {Sutton}},\ }\href {\doibase
  10.1103/PhysRevB.57.1505} {\bibfield  {journal} {\bibinfo  {journal} {Phys.
  Rev. B}\ }\textbf {\bibinfo {volume} {57}},\ \bibinfo {pages} {1505}
  (\bibinfo {year} {1998})}\BibitemShut {NoStop}%
\bibitem [{\citenamefont {Timrov}\ \emph {et~al.}(2022)\citenamefont {Timrov},
  \citenamefont {Marzari},\ and\ \citenamefont {Cococcioni}}]{2022TIM}%
  \BibitemOpen
  \bibfield  {author} {\bibinfo {author} {\bibfnamefont {I.}~\bibnamefont
  {Timrov}}, \bibinfo {author} {\bibfnamefont {N.}~\bibnamefont {Marzari}}, \
  and\ \bibinfo {author} {\bibfnamefont {M.}~\bibnamefont {Cococcioni}},\
  }\href@noop {} {\bibfield  {journal} {\bibinfo  {journal} {arXiv preprint
  arXiv:2203.15684}\ } (\bibinfo {year} {2022})}\BibitemShut {NoStop}%
\bibitem [{\citenamefont {Wang}\ \emph {et~al.}(2021)\citenamefont {Wang},
  \citenamefont {Liu}, \citenamefont {Jeon}, \citenamefont {Yi}, \citenamefont
  {Bang},\ and\ \citenamefont {Cho}}]{2021WAN}%
  \BibitemOpen
  \bibfield  {author} {\bibinfo {author} {\bibfnamefont {C.}~\bibnamefont
  {Wang}}, \bibinfo {author} {\bibfnamefont {S.}~\bibnamefont {Liu}}, \bibinfo
  {author} {\bibfnamefont {H.}~\bibnamefont {Jeon}}, \bibinfo {author}
  {\bibfnamefont {S.}~\bibnamefont {Yi}}, \bibinfo {author} {\bibfnamefont
  {Y.}~\bibnamefont {Bang}}, \ and\ \bibinfo {author} {\bibfnamefont {J.-H.}\
  \bibnamefont {Cho}},\ }\href {\doibase 10.1103/physrevb.104.l020504}
  {\bibfield  {journal} {\bibinfo  {journal} {Phys. Rev. B}\ }\textbf {\bibinfo
  {volume} {104}},\ \bibinfo {pages} {l020504} (\bibinfo {year}
  {2021})}\BibitemShut {NoStop}%
\bibitem [{\citenamefont {Chen}\ \emph {et~al.}(2012)\citenamefont {Chen},
  \citenamefont {Hu},\ and\ \citenamefont {Yang}}]{2012CHE}%
  \BibitemOpen
  \bibfield  {author} {\bibinfo {author} {\bibfnamefont {Y.}~\bibnamefont
  {Chen}}, \bibinfo {author} {\bibfnamefont {Q.-M.}\ \bibnamefont {Hu}}, \ and\
  \bibinfo {author} {\bibfnamefont {R.}~\bibnamefont {Yang}},\ }\href {\doibase
  10.1103/physrevlett.109.157004} {\bibfield  {journal} {\bibinfo  {journal}
  {Phys. Rev. Lett.}\ }\textbf {\bibinfo {volume} {109}},\ \bibinfo {pages}
  {157004} (\bibinfo {year} {2012})}\BibitemShut {NoStop}%
\bibitem [{\citenamefont {Vohra}\ \emph {et~al.}(1999)\citenamefont {Vohra},
  \citenamefont {Beaver}, \citenamefont {Akella}, \citenamefont {Ruddle},\ and\
  \citenamefont {Weir}}]{1999VOH}%
  \BibitemOpen
  \bibfield  {author} {\bibinfo {author} {\bibfnamefont {Y.~K.}\ \bibnamefont
  {Vohra}}, \bibinfo {author} {\bibfnamefont {S.~L.}\ \bibnamefont {Beaver}},
  \bibinfo {author} {\bibfnamefont {J.}~\bibnamefont {Akella}}, \bibinfo
  {author} {\bibfnamefont {C.~A.}\ \bibnamefont {Ruddle}}, \ and\ \bibinfo
  {author} {\bibfnamefont {S.~T.}\ \bibnamefont {Weir}},\ }\href {\doibase
  10.1063/1.369566} {\bibfield  {journal} {\bibinfo  {journal} {J. Appl.
  Phys.}\ }\textbf {\bibinfo {volume} {85}},\ \bibinfo {pages} {2451} (\bibinfo
  {year} {1999})}\BibitemShut {NoStop}%
\bibitem [{\citenamefont {Chen}\ \emph
  {et~al.}(2022{\natexlab{b}})\citenamefont {Chen}, \citenamefont {Liang},
  \citenamefont {Zhang}, \citenamefont {Song}, \citenamefont {Liu},
  \citenamefont {Jiang}, \citenamefont {Chen},\ and\ \citenamefont
  {Duan}}]{2022CHE}%
  \BibitemOpen
  \bibfield  {author} {\bibinfo {author} {\bibfnamefont {L.}~\bibnamefont
  {Chen}}, \bibinfo {author} {\bibfnamefont {T.}~\bibnamefont {Liang}},
  \bibinfo {author} {\bibfnamefont {Z.}~\bibnamefont {Zhang}}, \bibinfo
  {author} {\bibfnamefont {H.}~\bibnamefont {Song}}, \bibinfo {author}
  {\bibfnamefont {Z.}~\bibnamefont {Liu}}, \bibinfo {author} {\bibfnamefont
  {Q.}~\bibnamefont {Jiang}}, \bibinfo {author} {\bibfnamefont
  {Y.}~\bibnamefont {Chen}}, \ and\ \bibinfo {author} {\bibfnamefont
  {D.}~\bibnamefont {Duan}},\ }\href {\doibase 10.1088/1361-648x/ac55d7}
  {\bibfield  {journal} {\bibinfo  {journal} {J. Phys.: Condens. Matter}\
  }\textbf {\bibinfo {volume} {34}},\ \bibinfo {pages} {204005} (\bibinfo
  {year} {2022}{\natexlab{b}})}\BibitemShut {NoStop}%
\bibitem [{\citenamefont {Pickard}\ and\ \citenamefont
  {Needs}(2007)}]{2007PIC}%
  \BibitemOpen
  \bibfield  {author} {\bibinfo {author} {\bibfnamefont {C.~J.}\ \bibnamefont
  {Pickard}}\ and\ \bibinfo {author} {\bibfnamefont {R.~J.}\ \bibnamefont
  {Needs}},\ }\href {\doibase 10.1038/nphys625} {\bibfield  {journal} {\bibinfo
   {journal} {Nat. Phys.}\ }\textbf {\bibinfo {volume} {3}},\ \bibinfo {pages}
  {473} (\bibinfo {year} {2007})}\BibitemShut {NoStop}%
\bibitem [{\citenamefont {Akbarzadeh}\ \emph {et~al.}(2007)\citenamefont
  {Akbarzadeh}, \citenamefont {Ozoli{\c{n}}{\v{s}}},\ and\ \citenamefont
  {Wolverton}}]{2007RAK}%
  \BibitemOpen
  \bibfield  {author} {\bibinfo {author} {\bibfnamefont {A.~R.}\ \bibnamefont
  {Akbarzadeh}}, \bibinfo {author} {\bibfnamefont {V.}~\bibnamefont
  {Ozoli{\c{n}}{\v{s}}}}, \ and\ \bibinfo {author} {\bibfnamefont
  {C.}~\bibnamefont {Wolverton}},\ }\href {\doibase 10.1002/adma.200700843}
  {\bibfield  {journal} {\bibinfo  {journal} {Adv. Mater.}\ }\textbf {\bibinfo
  {volume} {19}},\ \bibinfo {pages} {3233} (\bibinfo {year}
  {2007})}\BibitemShut {NoStop}%
\bibitem [{\citenamefont {Virtanen}\ \emph {et~al.}(2020)\citenamefont
  {Virtanen}, \citenamefont {Gommers}, \citenamefont {Oliphant}, \citenamefont
  {Haberland}, \citenamefont {Reddy}, \citenamefont {Cournapeau}, \citenamefont
  {Burovski}, \citenamefont {Peterson}, \citenamefont {Weckesser},
  \citenamefont {Bright}, \citenamefont {van~der Walt}, \citenamefont {Brett},
  \citenamefont {Wilson}, \citenamefont {Millman}, \citenamefont {Mayorov},
  \citenamefont {Nelson}, \citenamefont {Jones}, \citenamefont {Kern},
  \citenamefont {Larson}, \citenamefont {Carey}, \citenamefont {Polat},
  \citenamefont {Feng}, \citenamefont {Moore}, \citenamefont {VanderPlas},
  \citenamefont {Laxalde}, \citenamefont {Perktold}, \citenamefont {Cimrman},
  \citenamefont {Henriksen}, \citenamefont {Quintero}, \citenamefont {Harris},
  \citenamefont {Archibald}, \citenamefont {Ribeiro}, \citenamefont
  {Pedregosa},\ and\ \citenamefont {van Mulbregt}}]{2020VIR}%
  \BibitemOpen
  \bibfield  {author} {\bibinfo {author} {\bibfnamefont {P.}~\bibnamefont
  {Virtanen}}, \bibinfo {author} {\bibfnamefont {R.}~\bibnamefont {Gommers}},
  \bibinfo {author} {\bibfnamefont {T.~E.}\ \bibnamefont {Oliphant}}, \bibinfo
  {author} {\bibfnamefont {M.}~\bibnamefont {Haberland}}, \bibinfo {author}
  {\bibfnamefont {T.}~\bibnamefont {Reddy}}, \bibinfo {author} {\bibfnamefont
  {D.}~\bibnamefont {Cournapeau}}, \bibinfo {author} {\bibfnamefont
  {E.}~\bibnamefont {Burovski}}, \bibinfo {author} {\bibfnamefont
  {P.}~\bibnamefont {Peterson}}, \bibinfo {author} {\bibfnamefont
  {W.}~\bibnamefont {Weckesser}}, \bibinfo {author} {\bibfnamefont
  {J.}~\bibnamefont {Bright}}, \bibinfo {author} {\bibfnamefont {S.~J.}\
  \bibnamefont {van~der Walt}}, \bibinfo {author} {\bibfnamefont
  {M.}~\bibnamefont {Brett}}, \bibinfo {author} {\bibfnamefont
  {J.}~\bibnamefont {Wilson}}, \bibinfo {author} {\bibfnamefont {K.~J.}\
  \bibnamefont {Millman}}, \bibinfo {author} {\bibfnamefont {N.}~\bibnamefont
  {Mayorov}}, \bibinfo {author} {\bibfnamefont {A.~R.~J.}\ \bibnamefont
  {Nelson}}, \bibinfo {author} {\bibfnamefont {E.}~\bibnamefont {Jones}},
  \bibinfo {author} {\bibfnamefont {R.}~\bibnamefont {Kern}}, \bibinfo {author}
  {\bibfnamefont {E.}~\bibnamefont {Larson}}, \bibinfo {author} {\bibfnamefont
  {C.~J.}\ \bibnamefont {Carey}}, \bibinfo {author} {\bibfnamefont
  {{\.{I}}.}~\bibnamefont {Polat}}, \bibinfo {author} {\bibfnamefont
  {Y.}~\bibnamefont {Feng}}, \bibinfo {author} {\bibfnamefont {E.~W.}\
  \bibnamefont {Moore}}, \bibinfo {author} {\bibfnamefont {J.}~\bibnamefont
  {VanderPlas}}, \bibinfo {author} {\bibfnamefont {D.}~\bibnamefont {Laxalde}},
  \bibinfo {author} {\bibfnamefont {J.}~\bibnamefont {Perktold}}, \bibinfo
  {author} {\bibfnamefont {R.}~\bibnamefont {Cimrman}}, \bibinfo {author}
  {\bibfnamefont {I.}~\bibnamefont {Henriksen}}, \bibinfo {author}
  {\bibfnamefont {E.~A.}\ \bibnamefont {Quintero}}, \bibinfo {author}
  {\bibfnamefont {C.~R.}\ \bibnamefont {Harris}}, \bibinfo {author}
  {\bibfnamefont {A.~M.}\ \bibnamefont {Archibald}}, \bibinfo {author}
  {\bibfnamefont {A.~H.}\ \bibnamefont {Ribeiro}}, \bibinfo {author}
  {\bibfnamefont {F.}~\bibnamefont {Pedregosa}}, \ and\ \bibinfo {author}
  {\bibfnamefont {P.}~\bibnamefont {van Mulbregt}},\ }\href {\doibase
  10.1038/s41592-020-0772-5} {\bibfield  {journal} {\bibinfo  {journal} {Nat.
  Methods}\ }\textbf {\bibinfo {volume} {17}},\ \bibinfo {pages} {352}
  (\bibinfo {year} {2020})}\BibitemShut {NoStop}%
\bibitem [{\citenamefont {Togo}\ and\ \citenamefont {Tanaka}(2015)}]{2015TOG}%
  \BibitemOpen
  \bibfield  {author} {\bibinfo {author} {\bibfnamefont {A.}~\bibnamefont
  {Togo}}\ and\ \bibinfo {author} {\bibfnamefont {I.}~\bibnamefont {Tanaka}},\
  }\href {\doibase 10.1016/j.scriptamat.2015.07.021} {\bibfield  {journal}
  {\bibinfo  {journal} {Scr. Mater.}\ }\textbf {\bibinfo {volume} {108}},\
  \bibinfo {pages} {1} (\bibinfo {year} {2015})}\BibitemShut {NoStop}%
\bibitem [{\citenamefont {Giannozzi}\ \emph {et~al.}(2009)\citenamefont
  {Giannozzi}, \citenamefont {Baroni}, \citenamefont {Bonini}, \citenamefont
  {Calandra}, \citenamefont {Car}, \citenamefont {Cavazzoni}, \citenamefont
  {Ceresoli}, \citenamefont {Chiarotti}, \citenamefont {Cococcioni},
  \citenamefont {Dabo}, \citenamefont {Corso}, \citenamefont {de~Gironcoli},
  \citenamefont {Fabris}, \citenamefont {Fratesi}, \citenamefont {Gebauer},
  \citenamefont {Gerstmann}, \citenamefont {Gougoussis}, \citenamefont
  {Kokalj}, \citenamefont {Lazzeri}, \citenamefont {Martin-Samos},
  \citenamefont {Marzari}, \citenamefont {Mauri}, \citenamefont {Mazzarello},
  \citenamefont {Paolini}, \citenamefont {Pasquarello}, \citenamefont
  {Paulatto}, \citenamefont {Sbraccia}, \citenamefont {Scandolo}, \citenamefont
  {Sclauzero}, \citenamefont {Seitsonen}, \citenamefont {Smogunov},
  \citenamefont {Umari},\ and\ \citenamefont
  {Wentzcovitch}}]{Giannozzi2009jcpm}%
  \BibitemOpen
  \bibfield  {author} {\bibinfo {author} {\bibfnamefont {P.}~\bibnamefont
  {Giannozzi}}, \bibinfo {author} {\bibfnamefont {S.}~\bibnamefont {Baroni}},
  \bibinfo {author} {\bibfnamefont {N.}~\bibnamefont {Bonini}}, \bibinfo
  {author} {\bibfnamefont {M.}~\bibnamefont {Calandra}}, \bibinfo {author}
  {\bibfnamefont {R.}~\bibnamefont {Car}}, \bibinfo {author} {\bibfnamefont
  {C.}~\bibnamefont {Cavazzoni}}, \bibinfo {author} {\bibfnamefont
  {D.}~\bibnamefont {Ceresoli}}, \bibinfo {author} {\bibfnamefont {G.~L.}\
  \bibnamefont {Chiarotti}}, \bibinfo {author} {\bibfnamefont {M.}~\bibnamefont
  {Cococcioni}}, \bibinfo {author} {\bibfnamefont {I.}~\bibnamefont {Dabo}},
  \bibinfo {author} {\bibfnamefont {A.~D.}\ \bibnamefont {Corso}}, \bibinfo
  {author} {\bibfnamefont {S.}~\bibnamefont {de~Gironcoli}}, \bibinfo {author}
  {\bibfnamefont {S.}~\bibnamefont {Fabris}}, \bibinfo {author} {\bibfnamefont
  {G.}~\bibnamefont {Fratesi}}, \bibinfo {author} {\bibfnamefont
  {R.}~\bibnamefont {Gebauer}}, \bibinfo {author} {\bibfnamefont
  {U.}~\bibnamefont {Gerstmann}}, \bibinfo {author} {\bibfnamefont
  {C.}~\bibnamefont {Gougoussis}}, \bibinfo {author} {\bibfnamefont
  {A.}~\bibnamefont {Kokalj}}, \bibinfo {author} {\bibfnamefont
  {M.}~\bibnamefont {Lazzeri}}, \bibinfo {author} {\bibfnamefont
  {L.}~\bibnamefont {Martin-Samos}}, \bibinfo {author} {\bibfnamefont
  {N.}~\bibnamefont {Marzari}}, \bibinfo {author} {\bibfnamefont
  {F.}~\bibnamefont {Mauri}}, \bibinfo {author} {\bibfnamefont
  {R.}~\bibnamefont {Mazzarello}}, \bibinfo {author} {\bibfnamefont
  {S.}~\bibnamefont {Paolini}}, \bibinfo {author} {\bibfnamefont
  {A.}~\bibnamefont {Pasquarello}}, \bibinfo {author} {\bibfnamefont
  {L.}~\bibnamefont {Paulatto}}, \bibinfo {author} {\bibfnamefont
  {C.}~\bibnamefont {Sbraccia}}, \bibinfo {author} {\bibfnamefont
  {S.}~\bibnamefont {Scandolo}}, \bibinfo {author} {\bibfnamefont
  {G.}~\bibnamefont {Sclauzero}}, \bibinfo {author} {\bibfnamefont {A.~P.}\
  \bibnamefont {Seitsonen}}, \bibinfo {author} {\bibfnamefont {A.}~\bibnamefont
  {Smogunov}}, \bibinfo {author} {\bibfnamefont {P.}~\bibnamefont {Umari}}, \
  and\ \bibinfo {author} {\bibfnamefont {R.~M.}\ \bibnamefont {Wentzcovitch}},\
  }\href {\doibase 10.1088/0953-8984/21/39/395502} {\bibfield  {journal}
  {\bibinfo  {journal} {J. Phys.: Condens. Matter}\ }\textbf {\bibinfo {volume}
  {21}},\ \bibinfo {pages} {395502} (\bibinfo {year} {2009})}\BibitemShut
  {NoStop}%
\bibitem [{\citenamefont {Corso}(2014)}]{2014DalCorso}%
  \BibitemOpen
  \bibfield  {author} {\bibinfo {author} {\bibfnamefont {A.~D.}\ \bibnamefont
  {Corso}},\ }\href {\doibase 10.1016/j.commatsci.2014.07.043} {\bibfield
  {journal} {\bibinfo  {journal} {Comp. Mater. Sci.}\ }\textbf {\bibinfo
  {volume} {95}},\ \bibinfo {pages} {337} (\bibinfo {year} {2014})}\BibitemShut
  {NoStop}%
\bibitem [{\citenamefont {McMillan}(1968)}]{1968MCM}%
  \BibitemOpen
  \bibfield  {author} {\bibinfo {author} {\bibfnamefont {W.~L.}\ \bibnamefont
  {McMillan}},\ }\href {\doibase 10.1103/physrev.167.331} {\bibfield  {journal}
  {\bibinfo  {journal} {Phys. Rev.}\ }\textbf {\bibinfo {volume} {167}},\
  \bibinfo {pages} {331} (\bibinfo {year} {1968})}\BibitemShut {NoStop}%
\bibitem [{\citenamefont {Allen}\ and\ \citenamefont {Dynes}(1975)}]{1975ALL}%
  \BibitemOpen
  \bibfield  {author} {\bibinfo {author} {\bibfnamefont {P.~B.}\ \bibnamefont
  {Allen}}\ and\ \bibinfo {author} {\bibfnamefont {R.~C.}\ \bibnamefont
  {Dynes}},\ }\href {\doibase 10.1103/physrevb.12.905} {\bibfield  {journal}
  {\bibinfo  {journal} {Phys. Rev. B}\ }\textbf {\bibinfo {volume} {12}},\
  \bibinfo {pages} {905} (\bibinfo {year} {1975})}\BibitemShut {NoStop}%
\bibitem [{\citenamefont {Liu}\ \emph {et~al.}(2017)\citenamefont {Liu},
  \citenamefont {Naumov}, \citenamefont {Hoffmann}, \citenamefont {Ashcroft},\
  and\ \citenamefont {Hemley}}]{2017LIU}%
  \BibitemOpen
  \bibfield  {author} {\bibinfo {author} {\bibfnamefont {H.}~\bibnamefont
  {Liu}}, \bibinfo {author} {\bibfnamefont {I.~I.}\ \bibnamefont {Naumov}},
  \bibinfo {author} {\bibfnamefont {R.}~\bibnamefont {Hoffmann}}, \bibinfo
  {author} {\bibfnamefont {N.~W.}\ \bibnamefont {Ashcroft}}, \ and\ \bibinfo
  {author} {\bibfnamefont {R.~J.}\ \bibnamefont {Hemley}},\ }\href {\doibase
  10.1073/pnas.1704505114} {\bibfield  {journal} {\bibinfo  {journal} {Proc.
  Natl. Acad. Sci.}\ }\textbf {\bibinfo {volume} {114}},\ \bibinfo {pages}
  {6990} (\bibinfo {year} {2017})}\BibitemShut {NoStop}%
\bibitem [{\citenamefont {Belli}\ \emph {et~al.}(2021)\citenamefont {Belli},
  \citenamefont {Novoa}, \citenamefont {Contreras-Garc{\'{\i}}a},\ and\
  \citenamefont {Errea}}]{2021BNCE}%
  \BibitemOpen
  \bibfield  {author} {\bibinfo {author} {\bibfnamefont {F.}~\bibnamefont
  {Belli}}, \bibinfo {author} {\bibfnamefont {T.}~\bibnamefont {Novoa}},
  \bibinfo {author} {\bibfnamefont {J.}~\bibnamefont
  {Contreras-Garc{\'{\i}}a}}, \ and\ \bibinfo {author} {\bibfnamefont
  {I.}~\bibnamefont {Errea}},\ }\href {\doibase 10.1038/s41467-021-25687-0}
  {\bibfield  {journal} {\bibinfo  {journal} {Nat. Commun.}\ }\textbf {\bibinfo
  {volume} {12}},\ \bibinfo {pages} {5381} (\bibinfo {year}
  {2021})}\BibitemShut {NoStop}%
\bibitem [{\citenamefont {Li}\ \emph {et~al.}(2019)\citenamefont {Li},
  \citenamefont {Miao}, \citenamefont {Ti}, \citenamefont {Liu}, \citenamefont
  {Chen}, \citenamefont {Shi},\ and\ \citenamefont {Gregoryanz}}]{2019LMT}%
  \BibitemOpen
  \bibfield  {author} {\bibinfo {author} {\bibfnamefont {B.}~\bibnamefont
  {Li}}, \bibinfo {author} {\bibfnamefont {Z.}~\bibnamefont {Miao}}, \bibinfo
  {author} {\bibfnamefont {L.}~\bibnamefont {Ti}}, \bibinfo {author}
  {\bibfnamefont {S.}~\bibnamefont {Liu}}, \bibinfo {author} {\bibfnamefont
  {J.}~\bibnamefont {Chen}}, \bibinfo {author} {\bibfnamefont {Z.}~\bibnamefont
  {Shi}}, \ and\ \bibinfo {author} {\bibfnamefont {E.}~\bibnamefont
  {Gregoryanz}},\ }\href {\doibase 10.1063/1.5130583} {\bibfield  {journal}
  {\bibinfo  {journal} {J. Appl. Phys.}\ }\textbf {\bibinfo {volume} {126}},\
  \bibinfo {pages} {235901} (\bibinfo {year} {2019})}\BibitemShut {NoStop}%
\bibitem [{\citenamefont {Bi}\ \emph {et~al.}(2022)\citenamefont {Bi},
  \citenamefont {Nakamoto}, \citenamefont {Shimizu}, \citenamefont {Zhou},
  \citenamefont {Wang}, \citenamefont {Liu},\ and\ \citenamefont
  {Ma}}]{2022BI}%
  \BibitemOpen
  \bibfield  {author} {\bibinfo {author} {\bibfnamefont {J.}~\bibnamefont
  {Bi}}, \bibinfo {author} {\bibfnamefont {Y.}~\bibnamefont {Nakamoto}},
  \bibinfo {author} {\bibfnamefont {K.}~\bibnamefont {Shimizu}}, \bibinfo
  {author} {\bibfnamefont {M.}~\bibnamefont {Zhou}}, \bibinfo {author}
  {\bibfnamefont {H.}~\bibnamefont {Wang}}, \bibinfo {author} {\bibfnamefont
  {G.}~\bibnamefont {Liu}}, \ and\ \bibinfo {author} {\bibfnamefont
  {Y.}~\bibnamefont {Ma}},\ }\href@noop {} {\bibfield  {journal} {\bibinfo
  {journal} {arXiv preprint arXiv:2204.04623}\ } (\bibinfo {year}
  {2022})}\BibitemShut {NoStop}%
\bibitem [{\citenamefont {Huang}\ \emph {et~al.}(2022)\citenamefont {Huang},
  \citenamefont {Luo}, \citenamefont {Dalladay~Simpson}, \citenamefont {Chen},
  \citenamefont {Cao}, \citenamefont {Peng}, \citenamefont {Gorelli},
  \citenamefont {Zhong}, \citenamefont {Lin},\ and\ \citenamefont
  {Chen}}]{2022HUA}%
  \BibitemOpen
  \bibfield  {author} {\bibinfo {author} {\bibfnamefont {G.}~\bibnamefont
  {Huang}}, \bibinfo {author} {\bibfnamefont {T.}~\bibnamefont {Luo}}, \bibinfo
  {author} {\bibfnamefont {P.}~\bibnamefont {Dalladay~Simpson}}, \bibinfo
  {author} {\bibfnamefont {L.}~\bibnamefont {Chen}}, \bibinfo {author}
  {\bibfnamefont {Z.}~\bibnamefont {Cao}}, \bibinfo {author} {\bibfnamefont
  {D.}~\bibnamefont {Peng}}, \bibinfo {author} {\bibfnamefont {F.~A.}\
  \bibnamefont {Gorelli}}, \bibinfo {author} {\bibfnamefont {G.}~\bibnamefont
  {Zhong}}, \bibinfo {author} {\bibfnamefont {H.}~\bibnamefont {Lin}}, \ and\
  \bibinfo {author} {\bibfnamefont {X.}~\bibnamefont {Chen}},\ }\href@noop {}
  {\bibfield  {journal} {\bibinfo  {journal} {arXiv preprint arXiv:2208.05199}\
  } (\bibinfo {year} {2022})}\BibitemShut {NoStop}%
\bibitem [{\citenamefont {Chen}\ \emph
  {et~al.}(2022{\natexlab{c}})\citenamefont {Chen}, \citenamefont {Luo},
  \citenamefont {Dalladay~Simpson}, \citenamefont {Huang}, \citenamefont {Cao},
  \citenamefont {Peng}, \citenamefont {Gorelli}, \citenamefont {Zhong},
  \citenamefont {Lin},\ and\ \citenamefont {Chen}}]{2022CHEc}%
  \BibitemOpen
  \bibfield  {author} {\bibinfo {author} {\bibfnamefont {L.}~\bibnamefont
  {Chen}}, \bibinfo {author} {\bibfnamefont {T.}~\bibnamefont {Luo}}, \bibinfo
  {author} {\bibfnamefont {P.}~\bibnamefont {Dalladay~Simpson}}, \bibinfo
  {author} {\bibfnamefont {G.}~\bibnamefont {Huang}}, \bibinfo {author}
  {\bibfnamefont {Z.}~\bibnamefont {Cao}}, \bibinfo {author} {\bibfnamefont
  {D.}~\bibnamefont {Peng}}, \bibinfo {author} {\bibfnamefont {F.~A.}\
  \bibnamefont {Gorelli}}, \bibinfo {author} {\bibfnamefont {G.}~\bibnamefont
  {Zhong}}, \bibinfo {author} {\bibfnamefont {H.}~\bibnamefont {Lin}}, \ and\
  \bibinfo {author} {\bibfnamefont {X.}~\bibnamefont {Chen}},\ }\href@noop {}
  {\bibfield  {journal} {\bibinfo  {journal} {arXiv preprint arXiv:2208.05191}\
  } (\bibinfo {year} {2022}{\natexlab{c}})}\BibitemShut {NoStop}%
\end{thebibliography}%

\clearpage
\newpage

\begin{table*}[htbp]
 \begin{center}
   \caption{
     Superconducting transition temperature ($T_\mathrm{c}$) predicted by the original McMillan formula (abbreviated to McM),~\cite{1968MCM} the Allen-Dynes-modified McMillan (McM) formula (abbreviated to AD),~\cite{1975ALL} and the model (i.e., Eq.~\eqref{eq:Tc_model}) proposed by Belli \textit{et al.},~\cite{2021BNCE} (abbreviated to ELF) for $A$-Ce-H ($A$ = Y and La) at some selected pressures. Accordingly, the values of some key parameters contained in these formulas and model are also listed.
   }
     \label{table.Pdep}
\begin{tabular}{lccccccccccc}
\hline
\hline
Phase &Space group & $P$  & N$_{atom}$& $d_{H-H}$&$\phi$ & $H_\mathrm{DOS}$ &$\lambda$ & $\omega_\mathrm{log}$&$T_\mathrm{c}$ (ELF)   &$T_\mathrm{c}$ (McM) &$T_\mathrm{c}$ (AD)  \\
 &   &(GPa)           &   & \AA &   &&  &(K) &(K) & (K) & (K) 	\\
 \hline
YCeH$_{8}$ & $P4/mmm$  & 100    & 20 & 1.52  & 0.40 & 9.5\%  & 0.41    & 987.5&24.5  & 4.7  &  4.8\\
YCeH$_{18}$ & $P\bar{6}m2$  & 150  & 22   &1.07& 0.43 &34.4\% & 2.51   	& 812.2 & 118.3 & 133.8  & 173.8 \\
YCeH$_{18}$ & $P\bar{6}m2$  & 300  & 22   &1.05& 0.46 &39.2\% & 0.97   	& 1547.7 & 142.2 & 103  & 110 \\
YCeH$_{20}$ & $R\bar{3}m$  & 300  & 66   &1.05& 0.45 &38.2\% & 1.03   & 1595.2 & 137.6 &114.2  & 122.1 \\
LaCeH$_{8}$ & $P4/mmm$  & 100   & 20   &1.56 &0.38 & 11.2\% & 0.35   & 975.8 & 24.9 & 1.73  &  1.76 \\
LaCeH$_{20}$ & $R\bar{3}m$ & 250  & 66  &1.06 &0.48 & 34.7\%	 &1.00 & 1566.9 &145.0  &109.1 & 116.1 \\
 \hline
 \hline
\end{tabular}
 \end{center}
\end{table*}


\end{document}